\documentclass[two column,showkeys]{revtex4}
\usepackage[utf8]{inputenc}
\usepackage[T1]{fontenc}
\usepackage{graphicx}
\usepackage{amsmath}
\usepackage{amssymb}
\usepackage{amsfonts}
\usepackage{color}
\usepackage{appendix}

\def\a{\alpha}
\def\b{\beta}
\def\g{\gamma}
\def\e{\varepsilon}
\def\d{\delta}

\def\l{\lambda}
\def\m{\mu}

\def\t{\tau}

\def\o{\omega}
\def\r{\rho}
\def\s{\sigma}
\def\z{\zeta}

\def\D{\Delta}

\def\be{\begin{equation}}
\def\ee{\end{equation}}
\def\bea{\begin{eqnarray}}
\def\eea{\end{eqnarray}}

\def\nn{\nonumber}
\def\lb{\label}

\def\XXint#1#2#3{{\setbox0=\hbox{$#1{#2#3}{\int}$}
     \vcenter{\hbox{$#2#3$}}\kern-.5\wd0}}

\begin{document}

\title{Impurity effects on Dirac modes in graphene armchair nanoribbons}

\author{Yuriy G. Pogorelov}%
	\email{ypogorel@fc.up.pt}
 	\affiliation{IFIMUP-IN,~Departamento~de~F\'{i}sica e Astronom\'{i}a,~Universidade~do~Porto,
 	~Porto,~Portugal,}
	
 	\author{Vadim M. Loktev}%
    \email{vloktev@bitp.kiev.ua}
 	\affiliation{N.~N.~Bogolyubov~Institute~of~Theoretical~Physics,~NAS~of~Ukraine,
 	~Kyiv,~Ukraine, \\
 	\&\\
 	Igor~Sikorsky~Kyiv~Polytechnic~Institute,~Kyiv,~Ukraine,}
 	
 \begin{abstract}
We consider finite ribbons of graphene with armchair orientation of their edges 
to study in detail impurity effects on specific Dirac-like modes. In the framework 
of Anderson hybrid model of impurity perturbation, a possibility for Mott localization 
and for opening of a mobility gap under local impurity perturbations is found and 
analyzed in function of this model parameters: the impurity energy level, its hybridization 
with the host Dirac modes, and the impurity concentration. Possible electronic phase states 
in such disordered system and subsequent phase transitions between them are discussed.
\end{abstract}

\date{\today}
\maketitle

\section{\label{sec:intr}Introduction}

Between electronic properties of two-dimensional (2D) graphene layer, the presence of linear gapless 
quasiparticle modes, or 2D Dirac modes, is especially notable by defining unusual physical effects 
in graphene \cite{Geim2004,Geim2005,Geim2009}. These modes are also a source for even finer, 1D Dirac 
modes, in graphene nanoribbons \cite{Yang,Yang2008,Ruffieux} with special orientation of their edges 
and special adjustment of their atomic width \cite{Wakabayashi1996}.

A broad family of Dirac semi-metals is of great interest for modern electronics, in 
particular, their behavior under doping by different impurities and the resulting 
restructuring of quasiparticle spectrum. Comparing with the known impurity effects 
in common semiconductors and in 2D graphene, the doped graphene nanoribbons can be 
expected to permit even higher sensitivity to various external controls and their 
study may deepen our general knowledge of disorder physics. 

This work continues the recent study of impurity effects in graphene nanoribbons 
\cite{Pogorelov2021}, focusing on their armchair edge orientation and their width 
adjusted for presence of Dirac-like modes in the electronic spectrum. In this course, 
we study various regimes of spectrum restructuring under impurity perturbation in 
function of perturbation parameters and compare the obtained results with the known 
such effects in other electronic materials.

The following consideration begins from the description of an armchair graphene nanoribbon (AGNR) 
and its spectral structure in terms of the second quantization Hamiltonian (Sec. \ref{Ham}) and 
the related Green functions (GFs, Sec. \ref{GF}). The perturbation of Hamiltonian by impurity adatoms 
within the Anderson hybrid model is introduced in Sec. \ref{Imp}, giving the solutions for perturbed 
GFs in the T-matrix form and checking for Mott localization of perturbed quasiparticles. The analysis 
of possible electronic phase states in doped AGNR in function of perturbation parameters is developed 
in Secs. \ref{Phase}, \ref{SO} and compared with the known behaviors of analogous electronic materials. 
The final discussion of the obtained results with suggestions for their possible practical use is 
presented in Sec. \ref{Dis}. An additional check of the solutions for 1D modes, beyond the T-matrix 
framework, is done in Appendix. 

It is our honour to dedicate this paper to 95th birthday of a prominent solid state theorist Emmanuel 
Iosifovich Rashba. His brilliant scientific career began in his native city Kyiv, Ukraine's capital, where 
he studied, worked for a long time and formed as a physicist under an outstanding supervise by S.I. Pecar. 
Rashba's results in non-ideal molecular crystal optics, in semiconductors theory, and, especially, in spintronics 
the very emergence of which was mostly due to his discovery of a new type spin-orbit coupling, essentially 
promoted the understanding of processes and phenomena in electronic systems of various dimensionalities.

\section{Hamiltonian and Green functions}\lb{Ham}
Graphene armchair nanoribbon can be seen as a periodic sequence of $N$ segments where each segment 
is a slant stack of $M$ layers (collinear between the segments) and each layer consists of two atomic 
sites (of graphene $a$- and $b$-types, see Fig. \ref{fig1}). 
\begin{figure}[h!]
    \centering 
    \includegraphics[width=8cm]{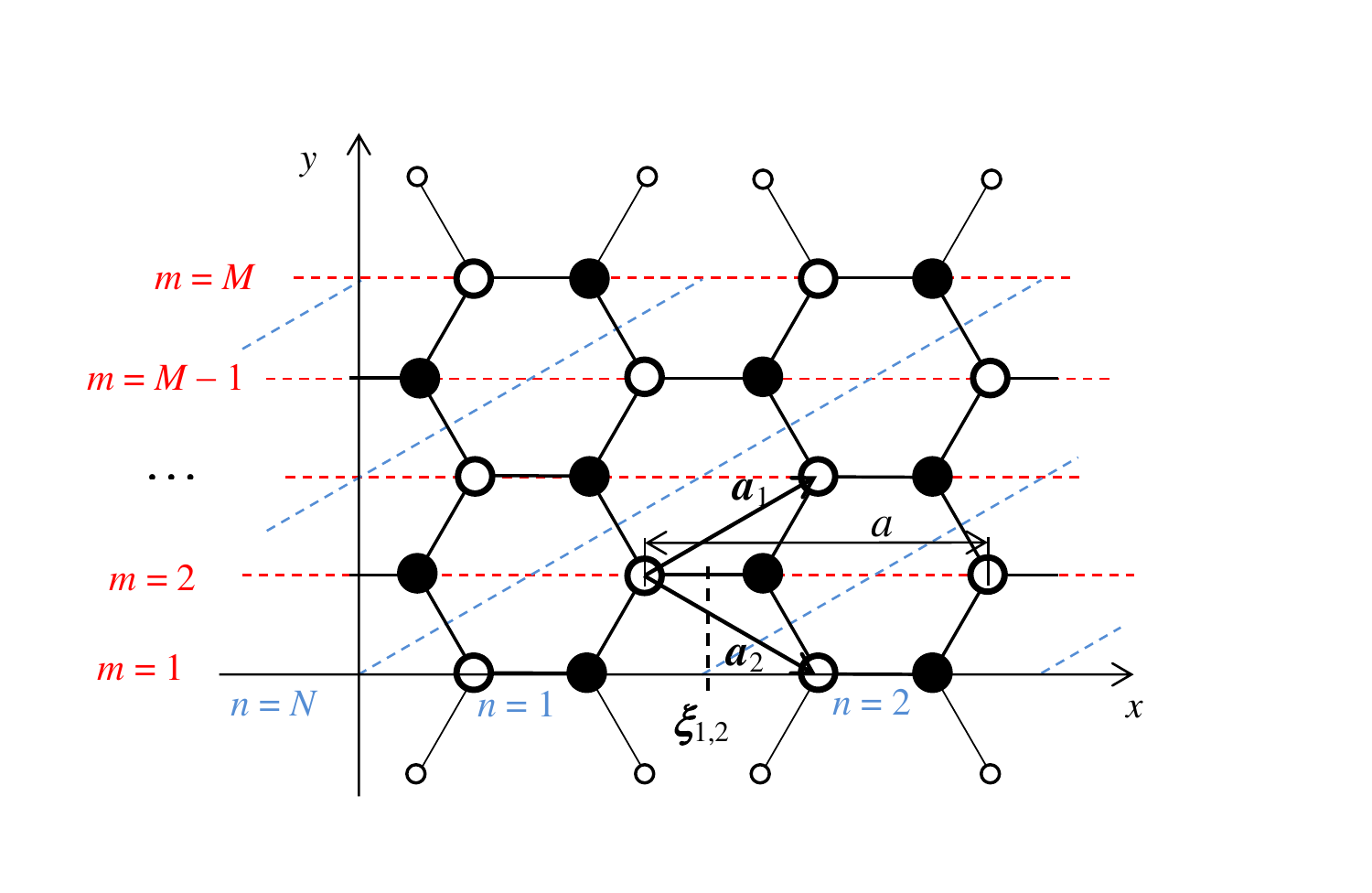}
\caption{Graphene nanoribbon with armchair orientation of its edges. Blue dashed lines delimit 
slant segments $n = 1,\dots,N$. Each segment extends along the graphene elementary translation 
vector ${\bf a}_1$ and includes $m = 1,\dots,M$ layers (red dashed lines) with $a$-type (white) 
and $b$-type (black) atomic sites. The sequence of segments has its longitudinal period $a = 
|{\bf a}_1 + {\bf a}_2|$. The carbon dangling bonds at the edges are passivated by hydrogens 
(small circles).}
\lb{fig1}
\end{figure} 
The respective local electronic states in $m$th layer of $n$th segment are generated by the local 
operators $a_{n,m}^\dagger$ and $b_{n,m}^\dagger$. Longitudinal translational invariance is imposed 
through the Born-von Karman closure of the $N$th to the 1st segment. For an AGNR with $M$ layers 
($M$-AGNR) and the nearest neighbor hopping $t$, the related tight-binding Hamiltonian reads:
\bea
H_{tb} & = & t\left\{\sum_{n=1}^N\left[\,\sum_{m=2}^M a_{n,m}^\dagger\left(b_{n,m} + b_{n-1,m+1}\right.
\right.\right.\nn\\
&& \qquad \left.\left. +\,\,b_{n,m-1}\right) + a_{n,1}^\dagger\left(b_{n,1} + b_{n-1,2}\right)
\right]\nn\\
 && \qquad\qquad + \left.\sum_{m=1}^{M-1} a_{1,m}^\dagger b_{N,m+1}  + h.c.
 \right\}.
\lb{H1}
\eea
The last sum in Eq.~\ref{H1} just generates the longitudinal translation invariance and suggests 
the Fourier-transform to 1D-wave operators. The longitudinal coordinates of $a$- and $b$-sites in 
units of the longitudinal period $a = |\mathbf{a}_1 + \mathbf{a}_2|$ define this transform as:
\bea
\a_{k,m} & = & \frac 1{\sqrt{N}}\sum_{n=1}^N {\rm e}^{i(2\pi k/N)(\xi_{n,m} - 1/6)}a_{n,m},\nn\\
\b_{k,m} & = & \frac 1{\sqrt{N}}\sum_{n=1}^N {\rm e}^{i(2\pi k/N)(\xi_{n,m} + 1/6)}b_{n,m},
\lb{Fta}
\eea
where $\xi_{n,m} = n + (m + 1)/2$ is the longitudinal coordinate of the center of $m$th layer from $n$th 
segment (see an example in Fig. \ref{fig1}). This readily diagonalizes the Hamiltonian, Eq. \ref{H1}, in 
the $k$ numbers. If the AGNR is macroscopically long, $N \to \infty$, one can pass to a quasi-continuous 
momentum variable: $2\pi k/N \to k$ (measured in $a^{-1}$ units). Also, for simplicity, the energy $\e$ 
will be measured in units of $t$. 

Then the system dynamics in the transversal $m$-index can be considered at a fixed longitudinal 
$k$-momentum, and the overall spectrum structure results qualitatively defined by the AGNR width 
$M$. In the known analytic approach by Wakabayashi {\it et al} \cite{Wakabayashi1996,Wakabayashi2009,Wakabayashi2010}, 
$2M$ eigen-states at given $k$ are taken as running $k$-waves superposed by standing waves in the transversal 
$q$-momentum, subject to the open edge condition (reaching a node when continued by a half-period beyond an 
AGNR edge). Namely, they are pairs of standing waves with discrete momentum values:
\be
q_j = \frac{\pi j}{M + 1}, \quad j = 1,\dots,M,
\lb{tm}
\ee
being just the combinations (symmetric and antisymmetric in $a$- and $b$-sites) of 1D-projected graphene 
states. 

The related eigen-energies are simple uniform 1D-projections of the 2D graphene spectrum for 
transversal momentum values $q_j$ by Eq. \ref{tm}: 
\be
\e_{j,k} = \sqrt{1 + 4\cos\tfrac k2\cos q_j + 4\cos^2q_j},
\lb{ev}
\ee
for conduction sub-bands (and $-\e_{j,k}$ for valence sub-bands). The 1D Brillouin zone (BZ) for all the $2M$ 
sub-bands is defined within the $0 \leq k \leq 2\pi$ range and the respective secular determinant reads:
\be
\det(\e - \hat H) = \prod_{j=1}^M \left(\e^2 - {\e_{j,k}}^2\right).
\lb{Am}
\ee

The eigen-state associated to the $(j,k)$-mode is a combination of the running $k$-wave and the 
standing $q_j$-wave \cite{Wakabayashi2010} with its amplitudes on $a$- and $b$-sites in $m$-layer:
\bea
&& A^{(j,k)}_m = -\frac{{\rm e}^{-i\varphi_{j,k}}}{\sqrt{M + 1}}\sin m q_j,\nn\\
&& \qquad\qquad B^{(j,k)}_m = \frac{{\rm e}^{i\varphi_{j,k}}}{\sqrt{M + 1}}\sin m q_j,
\lb{amp} 
\eea
where the phase is defined by the relation: 
\be
\varphi_{j,k} = \frac12\arctan\frac{\sin \tfrac{k}2 }{\cos\tfrac{k}2 + 2\cos q_j} + \frac k6.
\ee
The standing waves are orthonormalized through the relations:
\bea
\sum_{j = 1}^M \sin mq_j \sin m' q_j & = & \frac{M + 1}2\d_{m,m'},\nn\\
\sum_{m = 1}^M \sin mq_j \sin m q_{j'} & = & \frac{M + 1}2\d_{j,j'}.
\lb{ort}
\eea
Then we construct the eigen-mode operators $\psi_{\pm j,k}$ from the wave operators $\a_{m,k}$ and 
$\b_{m,k}$ by Eq. \ref{Fta} in order to reproduce the mode amplitudes by Eq. \ref{amp}: 
\bea
&& \psi_{\pm j,k} = \frac 1{\sqrt {M + 1}}\sum_{m = 1}^M \sin m q_j \left({\rm e}^{i\varphi_{j,k}}
\b_{m,k} \right.\nn\\
&& \qquad\qquad\qquad\qquad\qquad\qquad \mp \left. {\rm e}^{-i\varphi_{j,k}}\a_{m,k}\right).
\lb{rel}
\eea
In their basis, the Hamiltonian, Eq. \ref{H1}, turns fully diagonal:
\be
H_{tb} = \sum_{j,k}\e_{j,k}\left(\psi_{j,k}^\dagger \psi_{j,k} - \psi_{-j,k}^\dagger \psi_{-j,k}\right).
\lb{Hd}
\ee

By inversion of Eqs. \ref{Fta}, \ref{rel}, the local operators are expanded in the eigen-mode operators:
\bea
&&a_{n,m} = \frac 1{\sqrt{(M + 1)N}}\sum_{j,k}{\rm e}^{i(k\xi_{n,m} - \varphi_{j,k})}\sin m q_j\nn\\
&&\qquad\qquad\qquad\qquad \times\,\left(\psi_{-j,k} - \psi_{j,k}\right),\nn\\
&&b_{n,m} = \frac 1{\sqrt{(M + 1)N}}\sum_{j,k}{\rm e}^{i(k\xi_{n,m} + \varphi_{j,k})}\sin m q_j\nn\\
&&\qquad\qquad\qquad\qquad \times\,\left(\psi_{-j,k} + \psi_{j,k}\right),
\lb{rel1}
\eea
which is helpful for the next treatment of AGNR perturbations by local impurity centers.  
 
\begin{figure}[h]
        \includegraphics[width=8cm]{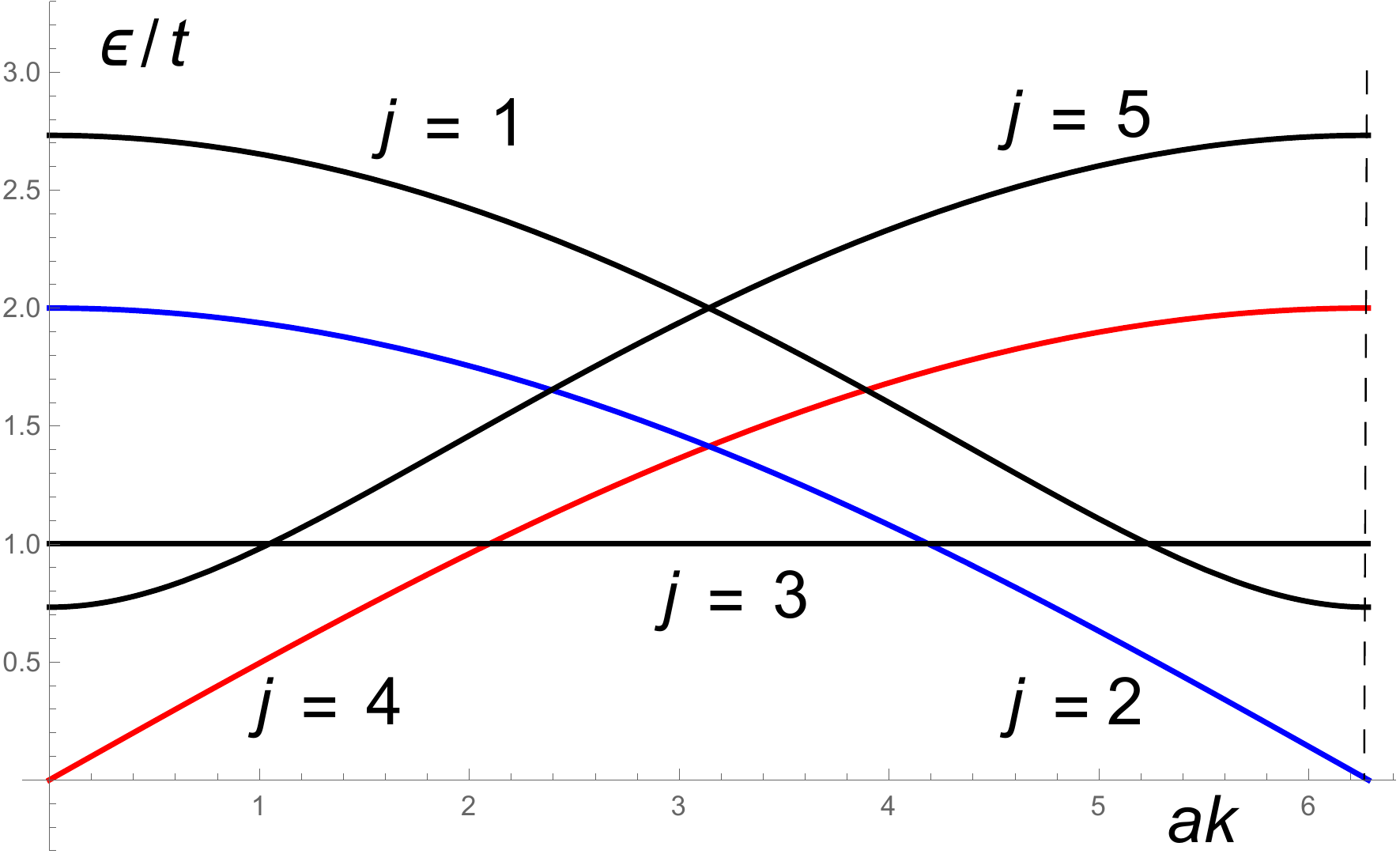}
    \caption{Energy bands dispersion in 5-AGNR ($\nu = 2$ by Eq. \ref{mod3}), showing the 
    Dirac-like modes with $j = 2,4$.}
    \label{fig2}
\end{figure}

The notable feature of the spectrum by Eq. \ref{ev} is that it contains gapless modes if the AGNR width 
satisfies a special condition \cite{Wakabayashi1996}: 
\be
M + 1 = 3\nu,\quad \nu = 1,2,\dots.
\lb{mod3}
\ee
For such $M = 3\nu - 1$, the mode with $j = 2\nu$ reaches zero energy at the BZ edge $k = 0$ 
as:
\be
\e_{2\nu,k} = 2\left|\sin \frac { k}4\right| \approx \frac{|k|}2
\lb{A0}
\ee 
(see Fig. \ref{fig2}), and the mode with $j = \nu$ reaches zero energy at the opposite BZ edge 
$k = 2\pi$ as: 
\be
\e_{\nu,k} = 2\left|\cos \frac{k}4\right| \approx \frac{|k- 2\pi|}2.
\lb{A1}
\ee
The dispersion laws by Eqs. \ref{A0}, \ref{A1} formally coincide with the standard linear 
dispersion near the Dirac points of 2D graphene, hence they can be seen as definitions of 
effective 1D Dirac points in ($3\nu -1$)-AGNR spectra. All other modes there (with $j \neq \nu, 2\nu$) 
have finite energy gaps. 

\section{Green functions and observables}\lb{GF}
The following consideration goes in the framework of two-time GFs \cite{Zubarev1960,
Economou1979} defined by their Fourier transforms:
\be
\langle\langle A|B\rangle\rangle_\e = \frac i\pi \int_0^\infty dt{\rm e}^{it(\e + i0)} 
\left\langle \left\{A(t),B(0)\right\}\right\rangle
\lb{ttGF}
\ee
where $A(t) = {\rm e}^{i Ht}A{\rm e}^{-i Ht}$ is a Heisenberg picture operator for the system 
Hamiltonian $H$, $\{.,.\}$ is the anti-commutator and $\langle\dots\rangle$ is the 
quantum-statistical average. In what follows, the GF's energy subindex is mostly omitted. 

Practical calculation of GFs is done through the general equation of motion:
\be
\e\langle\langle A|B\rangle\rangle =  \langle \left\{A(0),B(0)\right\}\rangle + 
\langle\langle \left[A,H\right]|B\rangle\rangle,
\lb{eqmot}
\ee
involving the commutator $\left[.,.\right]$. So found GFs generate the physical observable 
quantities (the averages of operator products) through the spectral relation:
\be 
\langle AB\rangle = \frac 1\pi {\rm Im} \int_{0}^{\infty}\langle\langle B|A\rangle
\rangle_\e d\e.
\lb{spec}
\ee
In the present case, the system electronic properties can be obtained from the 2M$\times$2M 
GF matrix $\hat G(k,k')$ with its matrix elements $G_{j,j'}(k,k') \equiv \langle\langle \psi_{j,k}
|\psi_{j',k'}^\dagger\rangle\rangle$ built from the eigen-mode operators by Eq. \ref{rel} where the 
$j$-indices count the transversal momenta as by Eq. \ref{tm} and also their opposites $-j$ ($2M$ 
altogether). 
 
For the unperturbed AGNR system with its diagonal Hamiltonian, Eq. \ref{Hd}, the above defined 
GF matrix results also diagonal: $G_{j,j'}^{(0)}(k,k') = \d_{j,j'}\d_{k,k'}G_{j,k}^{(0)}(\e)$, 
with its diagonal elements called propagators:
\be
G_{j,k}^{(0)}(\e) = \frac 1{\e - \e_{j,k}}.
\lb{gfa}
\ee
They define an important observable, the density of quasiparticle states (DOS), as a sum $\r(\e) 
= \sum_{j = 1}^M \r_j(\e)$ where a partial contribution from $(j,-j)$ subbands is:
\be
\r_j(\e) = \frac 2\pi\sum_k {\rm Im}\,\left[G_{j,k}^{(0)}(\e) +  G_{-j,k}^{(0)}(\e)\right]
\lb{ds}
\ee
(including the implicit factor 2 for electron spins). Using Eqs. \ref{ev}, \ref{gfa} and passing from sum in 
$k$ to integral:
\be
 \frac 1N \sum_k f(k) = \frac 1{2\pi} \int_0^{2\pi}f(k) dk,
 \lb{int}
 \ee  
gives this contribution as:
\be
 \r_j(\e) = \frac{8|\e|}{\pi (M + 1)\sqrt{(\e^2 - \e_{-,j}^2)(\e_{+,j}^2 - \e^2)}},
 \lb{ds1}
 \ee
where $\e_{\pm,j} = 1 \pm 2\cos q_j$ are the $j$th subband energy limits. 
\begin{figure}[h]
        \includegraphics[width=8cm]{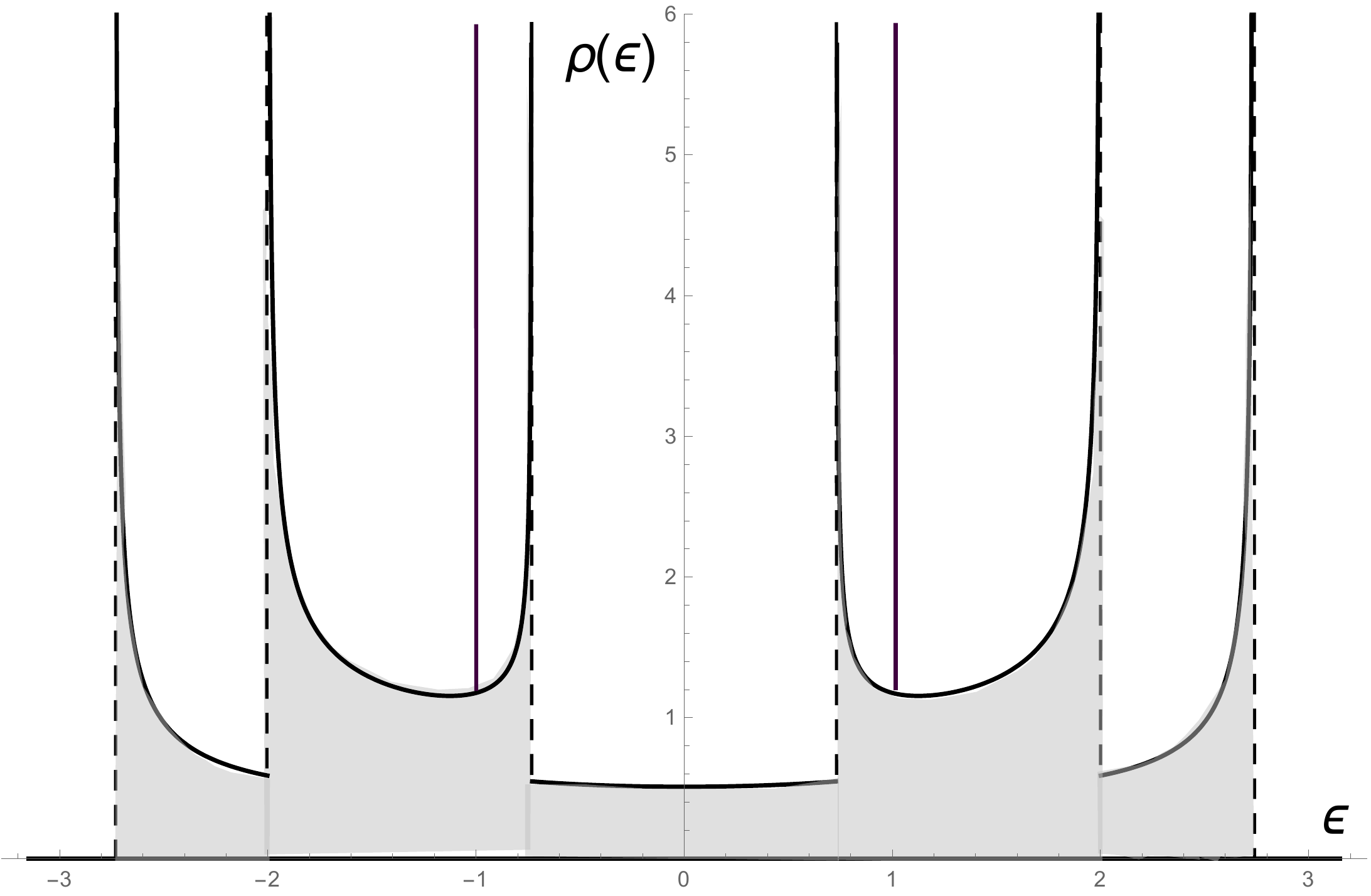}
    \caption{Density of states in 5-AGNR. Almost constant value $\approx 4/(3\pi)$ within the 
    low energy range, $|\e| \leq \sqrt3 - 1$, comes from the Dirac-like ($j = 2,4$) subbands.}
    \label{fig3}
\end{figure}

In particular, for the most relevant here Dirac-like modes with $\e_{+,\nu} = \e_{-,2\nu} = 2,\,
\e_{+,2\nu} = \e_{-,\nu} = 0$, we have:
 \be
 \r_\nu(\e) = \r_{2\nu}(\e) = \frac 4{3\pi \nu \sqrt{1 - (\e/2)^2}},
 \lb{rd}
 \ee
so DOS results almost constant at low energies: $\r(\e) \approx \r(0) \equiv \r_0 = 8/(3\pi \nu)$ 
at $|\e| \ll 1$, as seen in the case of 5-AGNR presented in Fig. \ref{fig3}.

\section{Impurity perturbations}\lb{Imp}

The local impurity perturbations in graphene nanoribbons of both zigzag and armchair types were 
recently considered within two basic impurity models: the one-parameter Lifshitz model (LM) 
\cite{Lifshitz_1963}, more adequate for substitutional impurities, and the two-parameter Anderson 
hybrid model (AM) \cite{Anderson}, for impurity adatoms \cite{Pogorelov2021}. For zigzag structures, 
the overall conclusion was about their eigen-modes stability (topological protection) against 
quasiparticles localization by the impurity disorder, both in LM and AM models.

However such localization was found in AGNRs with LM impurities, though reduced in that case to a 
narrow vicinity of the Dirac point (zero energy). But this already opens a possibility for Mott's 
metal/insulator phase transitions in a nanosystem and generates the next interest for studying AGNR 
behavior under more diversified AM perturbations. In the latter case, a more complicated intermittence 
of conducting and localized states in other ranges of energy spectrum and a broader variety of related 
phase states for this 1D system can be expected. Then it would be also of interest to compare such 
effects with the known analogs for 3D and 2D electronic systems under impurity disorder.

The following consideration is focused on special ($3\nu - 1$)-AGNRs and restricted to only their 
Dirac-like modes. Since these modes with $j = \nu$ and $2\nu$ give identical and independent contributions 
to the spectrum, one can next focus on a single Dirac-like mode, say $j = 2\nu$, then denoting $\psi_{2\nu,k} 
\equiv \psi_{+,k}$ and $\psi_{-2\nu,k} \equiv \psi_{-,k}$. Consequently, the above introduced $\hat G(k,k')$ 
matrix gets reduced to the 2$\times$2 form in the basis of $\psi_{\pm,k}$ operators. In particular, the 
non-perturbed solution, Eq. \ref{gfa}, presents here as $\hat G^{(0)}(k,k') = \d_{k,k'}(\e - \e_k\hat
\s_3)^{-1}$ with $\e_k \equiv \e_{2\nu,k}$ by Eq. \ref{A0} and the Pauli matrix $\hat\s_3$ in $\pm$ 
indices. 
\begin{figure}[h]
\includegraphics[width=9cm]{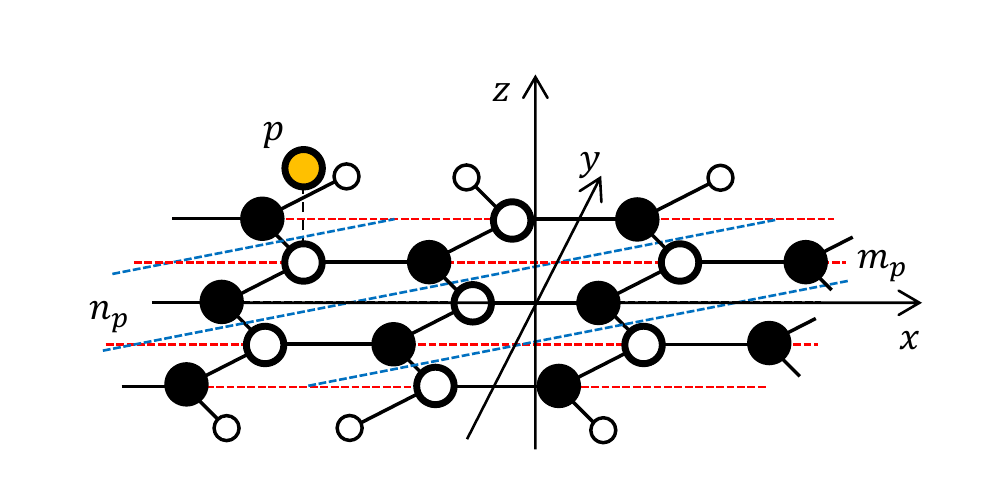}
\caption{Position $p$ of a top impurity adatom over an $a$-site in $m_p$-layer from $n_p$-segment 
of 5-AGNR.}
\lb{fig4}
\end{figure}

Then we consider the impurity adatoms location restricted to the simplest top-positions: $p_a$ over 
an $a$-type host atom in $m_p$ layer from $n_p$ segment (see Fig. \ref{fig4}) or $p_b$ over a $b$-type 
host atom. 

In these notations, the AM perturbation Hamiltonian reads:
\bea
H_{AM} & = & \sum_p\left\{\e_{res}c_p^\dagger c_p + \frac{\o}{\sqrt{3\nu N}}\sum_k\left[\sin\frac{\pi m_p}3
\right.\right.\nn\\
& \times & \left.\left.{\rm e}^{i(k\xi_p \mp \phi_k)}c_p^\dagger(\psi_{-,k} \mp \psi_{+,k}) + {\rm h.c.}
\right]\right\},
\lb{ArmL}
\eea
where a local impurity operator $c_p$ with its resonance level $\e_{res}$ is coupled to the Dirac-like 
modes $\pm\e_k$ through the hybridization $\o$ with the neighbor host atom at the longitudinal position 
$\xi_{n_p,m_p} - 1/6$ for its $a$-type as in Fig. \ref{fig4} (or $\xi_{n_p,m_p} + 1/6$ for its $b$-type).

Then the complete Hamiltonian $H_{tb} + H_{AM}$ generates a perturbation of the GF matrix: $\hat G^{(0)} 
\to \hat G$. In its simplest form, this is given by the T-matrix approximation:
\be
\hat G = \left(\e - c\hat T(\e) - \e_k\hat\s_3\right)^{-1},
\lb{TM}
\ee
where $c = (2MN)^{-1}\sum_p 1$ is the impurity concentration and the T-matrix in this case 
results diagonal:
\be
\hat T(\e) \equiv T(\e) = \frac{\o^2}2\left(\e - \e_{res} - \frac{i\o^2}{4f\sqrt{1 - 
(\e/2)^2}}\right)^{-1}.
\lb{TM1}
\ee
\begin{figure}[h!]
        \includegraphics[width=8cm]{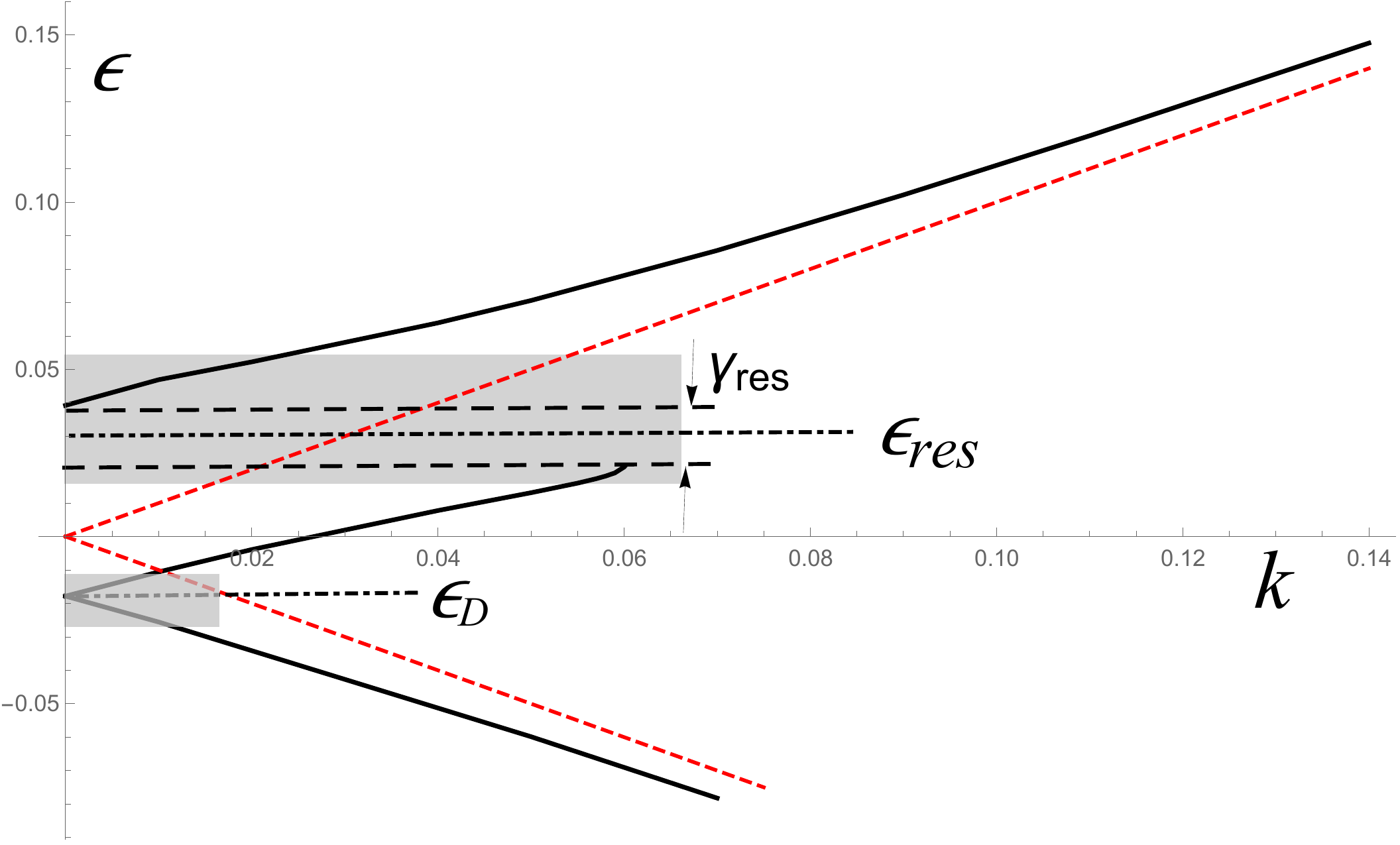}
\caption{Dispersion law for the Dirac-like modes in 5-AGNR having AM impurities with their
parameters $\e_{res} = 0.03$, $\o = 0.3$, and concentration $c = 0.02$. For comparison, the 
unperturbed Dirac-like modes are shown by the red dashed lines.}
\lb{fig5}
\end{figure}

Then the modified dispersion law $\tilde\e_k$ follows from the standard GF secular equation 
${\rm Re}\,[\det (\hat {G})^{-1}] = 0$ \cite{Bonch} as: 
\be
 \tilde\e_k = \sqrt{\e_k^2 + (c\,{\rm Im}\,T(\e))^2} + c\,{\rm Re}\,T(\e),
\lb{ek}
\ee 
and its solution for 5-AGNR at the choice of AM impurity parameters $\e_0 = 0.03$ and $\o = 0.3$, 
corresponding to Cu adatoms in top position \cite{Pogorelov2020}, is presented in Fig. \ref{fig5}. 
Here the characteristic impurity effects are seen as:

1) the shift of the Dirac point from its initial zero energy position down to
\be
\e_{\rm D} \equiv \tilde\e_{k = 0} \approx \frac{\e_{res} -\sqrt{\e_{res}^2 + 2 c\o^2}}2,
\lb{ed}
\ee

2) the resonance splitting between the initial linear $\e_k$ law and the impurity resonance level 
$\e_{res}$ until its vicinity of width 
\be
\g_{res} \approx \o\sqrt{\frac {c - c_0}2}.
\lb{d0}
\ee
where $c_0 \sim \o^2/(8\nu^2)$ is the critical concentration value for this splitting to appear.

Also an anomalous negative dispersion formally appears inside this vicinity, at $|\e -\e_{res}| 
\lesssim \g_{res}$, but this range occurs unphysical when validity of the modified dispersion law is 
checked with the Ioffe-Regel-Mott (IRM) criterion for conducting states \cite{IoffeRegel,
Mott}:
\be
k\frac{\partial\tilde\e_k}{\partial k} \gtrsim c\,{\rm Im}\,T(\tilde\e_k).
\lb{IRM}
\ee

This simply means that the quasiparticle lifetime (inverse of the r.h.s) is longer than its oscillation 
period (inverse of the l.h.s) and such quasiparticles are conductive indeed, otherwise they are localized 
near impurity sites. The explicit form of the IRM criterion for the dispersion law by Eq. \ref{ek} and 
the T-matrix by Eq. \ref{TM1} reads:
\be
\frac 2{\partial\ln{\rm Re}\,\left[\e - cT(\e)\right]^2/\partial\e}\gtrsim c\,{\rm Im}\,T,
\lb{IRM1}
\ee
All the energy ranges where this inequality does not hold are attributed to localized states so the 
dispersion law by Eq. \ref{ek} for conducting states does not apply there. The mobility edges between 
conducting and localized states can be qualitatively estimated as the $\e$ values that make the relation 
of Eq. \ref{IRM1} an equality. The results of such numerical estimation at the choice of AM parameters 
as for Cu top impurities are shown in Fig. \ref{fig6}a. They illustrate formation of two mobility gaps 
(ranges of localization), one around $\e_{res}$ and another around $\e_D$. Their width grows with the 
impurity concentration $c$: the first as $\sim 2\g_{res}(c)$ (by Eq. \ref{d0}) and the second as $\g_D 
\approx c\o^4/[8\nu(\e_D - \e_{res})^2]$. 
\begin{figure}[h]
        \includegraphics[width=8cm]{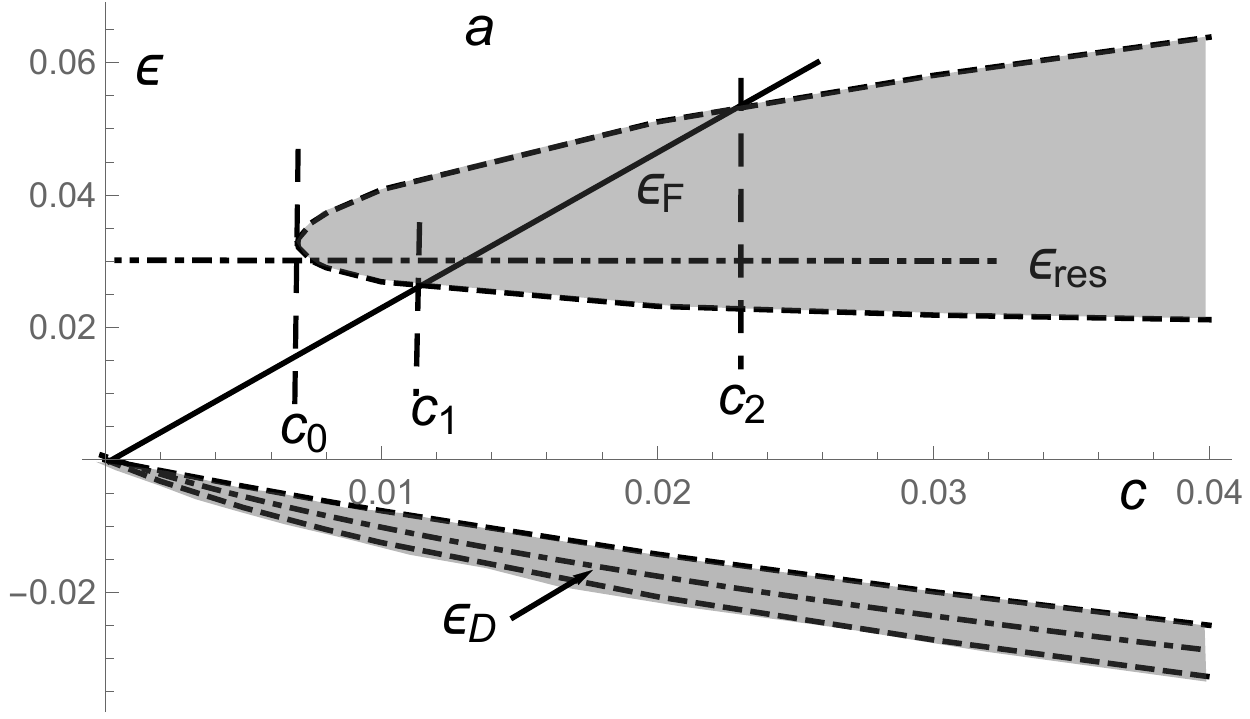}
         \includegraphics[width=8cm]{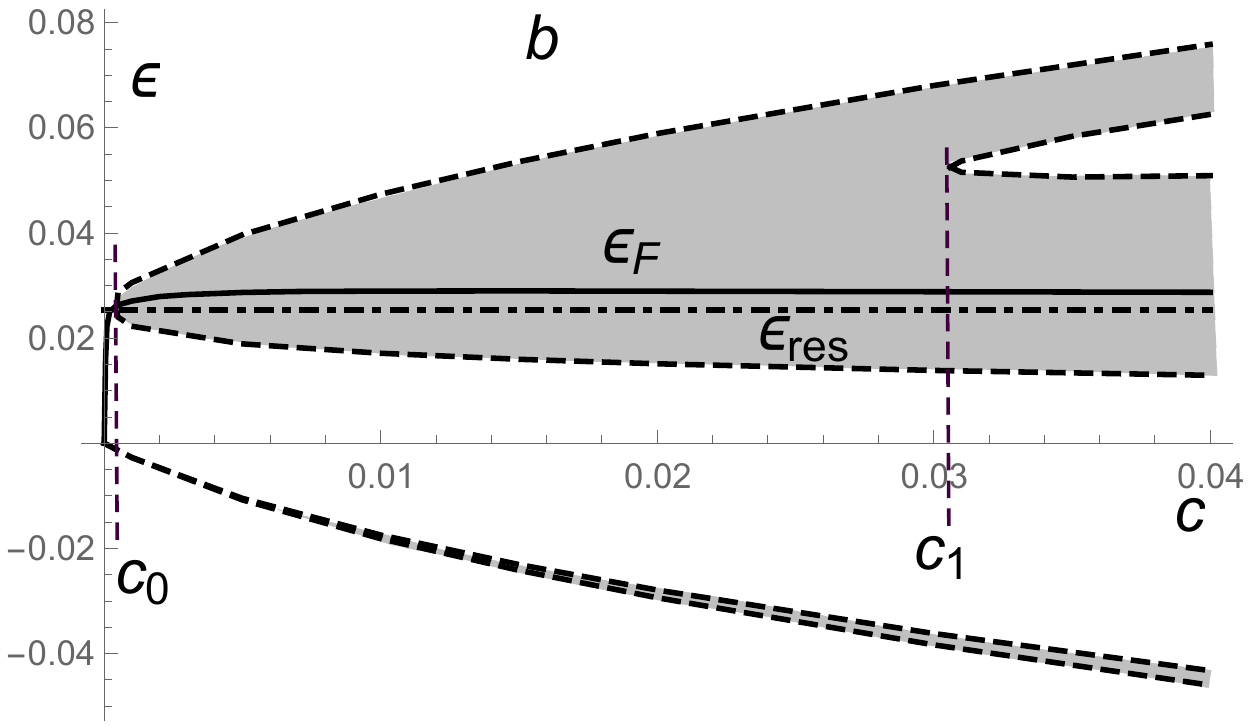}
\caption{a) Mobility gaps (in meV scale) in function of impurity concentration $c$ at 
the choice of AM parameters as in Fig. \ref{fig5}. Special concentration values refer respectively: 
$c_0$ to opening of mobility gap near the resonance level $\e_{res}$, $c_1$ to metal/insulator, 
and $c_2$ to insulator/metal transitions (see below). b) Analogous developments for the 
same impurities in 2D graphene, note here opening of a spectrum quasi-gap inside the mobility gap 
at $c_1$, absent in AGNR.}
\lb{fig6}
\end{figure}

Validity of the above T-matrix description is confirmed with an additional check beyond the frame 
of this simplest approximation (see in Appendix).

An important issue for such disordered AGNRs is how the positions of mobility edges compare with 
that of the Fermi level {\it vs} impurity concentration, $\e_{\rm F}(c)$. The latter results from 
an extra amount of $c$ charge carriers by impurities (per host atom) being filled into the relevant 
subbands: 
\be
\int_0^{\e_{\rm F}(c)}\r(\e)d\e = c.
\lb{ef}
\ee
Notably, the above indicated low energy DOS, $\r(\e) \approx \r_0$, holds its constancy even under 
impurity disorder, as seen from stability of Eq. \ref{rd} at the passing $\e \to \e - cT(\e)$ since $|\e - cT(\e)
| \ll 1$ for all $|\e| \ll 1$, one of the main specifics of this 1D-like system. 

Then from Eq. \ref{ef} we come to a simple linear relation $\e_{\rm F}(c) \approx c/\r_0$. Such behavior 
superposed onto the diagram of mobility gaps in Fig. \ref{fig6}a shows a possibility for $\e_{\rm F}$ 
to cross the mobility edges near $\e_{res}$, both into and out of this mobility gap. Then an 
intermittency of the related metal/insulator transitions (MIT) can be expected.

It is of interest to compare these AGNR results with their analogs for the same impurities in 2D 
graphene (Fig. \ref{fig6}b). Besides a general similarity of two pictures, they also present substantial 
differences. First of all, the estimated critical concentration for localization on Cu impurities in 5-AGNR, 
$c_0 \approx 7\cdot 10^{-3}$, is more than an order of magnitude higher than its analog for the 2D 
case, $c_0^{(2D)} \approx 4\cdot 10^{-4}$ \cite{Pogorelov2020}. This can be explained by the higher 
and almost constant low-energy DOS for the 1D Dirac-like modes: $\r_0 \sim 1$, compared to its linear 
in $\e$ smallness for the 2D Dirac modes: $\r_{2D}(\e) \approx 4\e/(\pi\sqrt 3) \ll 1$. 

Another difference is in behaviors of $\e_{\rm F}(c)$ for each system, also caused by that of DOS. 
From Eq. \ref{ef}, its almost linear growth in AGNR: $\e_{\rm F}(c) \approx c/\r_0$ is much slower 
than the very fast, square root initial growth of $\e_{\rm F}^{(2D)}(c) \approx \sqrt{c\pi\sqrt3/2}$ in 2D 
graphene. This defines a much higher threshold in $c$ for occurrence of MIT in AGNR: $c_1 \sim \e_{res}
\r_0$, than in 2D graphene: $\sim c_0^{(2D)}$. But the constancy of AGNR DOS, even in presence of 
impurities, permits the linear $\e_{\rm F}(c)$ growth to persist also for $c > c_1$ and then an inverse 
MIT to occur at its emergence from the mobility gap at $c = c_2 \sim \e_{res}\r_0 + (\o\r_0)^2$. In 
contrary, the 2D graphene DOS presents a sharp resonance peak near $\e_{res}$ and the Fermi level, 
when approaching this peak at $c \to c_0^{(2D)}$, gets fixed here, defining insulating phase for 
all $c > c_0^{(2D)}$.

At least, the critical concentration $c_0$ by Eq. \ref{d0} that defines the onset of localization 
near $\e_{res}$, indicates it to occur earlier for weaker impurity-host coupling $\o$.

\section{Electronic phase states and their tuning}\lb{Phase}
An important physical issue is to determine the system electronic phase states. For the considered 
nanoribbons, this refers first of all to their electric conductivity. 

In the limit of zero temperature, it is fully defined by the Fermi level position with respect to the spectrum 
mobility edges: implying the metallic state for $\e_{\rm F}$ out of the mobility gaps and the insulating state 
for $\e_{\rm F}$ inside them. Thus, from the diagram in Fig. \ref{fig6}, the system of 5-AGNR with given 
concentration $c$ of Cu impurities is expected to be metallic if $0 < c < c_1$ or $c > c_2$ and insulating 
if $c_1 < c < c_2$.

But a practical interest arises in an effective tuning of possible MIT's at a given impurity composition (in 
analogy with the common gate controls in doped semiconductors). First of all, the initial composition can 
be chosen to set the Fermi level close enough to a mobility edge, for instance, $\e_{\rm F} < \e_g < 
\e_{res}$ and $\e_{\rm F} - \e_g \ll \e_{res}$. Then several factors can be considered for MIT tuning: 
1) temperature (in terms of its inverse $\b$), 2) magnetic field, $\bf h$, and 3) electric field, $\bf E$. 

The temperature control will result from the interplay between the metallic conductivity $\s_{met}
(\b)$, due to extended states, and the hopping conductivity $\s_{hop}(\b)$, due to localized states. 

The first type refers to the Kubo-Greenwood formula written here as:
\be
\s_{met}(\b) \approx \r_0\int_0^{\e_g}[1 - (\e/2)^2]\t(\e)\frac{\partial n(\e,\b)}{\partial\e}d\e,
\lb{kubo}
\ee
 where $n(\e,\b)  = [{\rm e}^{\b(\e - \e_{\rm F})}  + 1]^{-1}$  is the standard Fermi function and 
the lifetime $\t(\e) = [\t_{imp}^{-1}(\e) + \t_{ph}^{-1}(\e)]^{-1}$ involves: 

$\t_{imp}^{-1}(\e) = c|{\rm  Im}\,T(\e)|$ ($\b$-independent) due to impurities, and 

$\t_{ph}^{-1}(\e)  \sim 1/\b\Theta_{\rm D}$, due to 1D phonons (with  the Debye temperature $\Theta_{\rm D}$). 

The second type refers to the Mott formula written for an 1D system:
\be
\s_{hop}(\b) \propto {\rm e}^{-\sqrt{\b T_0}},
\lb{mott}
\ee
where $T_0 \sim \t^{-1}(\e_{\rm F})/\r(\e_{\rm F})$. Interplay of this decreasing $\s_{hop}(\b)$ and the growing 
$\s_{met}(\b)$ by Eq. \ref{kubo} results in an overall  conductivity maximum at $\b \sim 1/|\e_{\rm F} - \e_g|$, 
but having a comparable temperature width.  So, this crossover between the types of conductivity is not yet a 
canonical phase transition.

\begin{figure}[h]
\includegraphics[width=8cm]{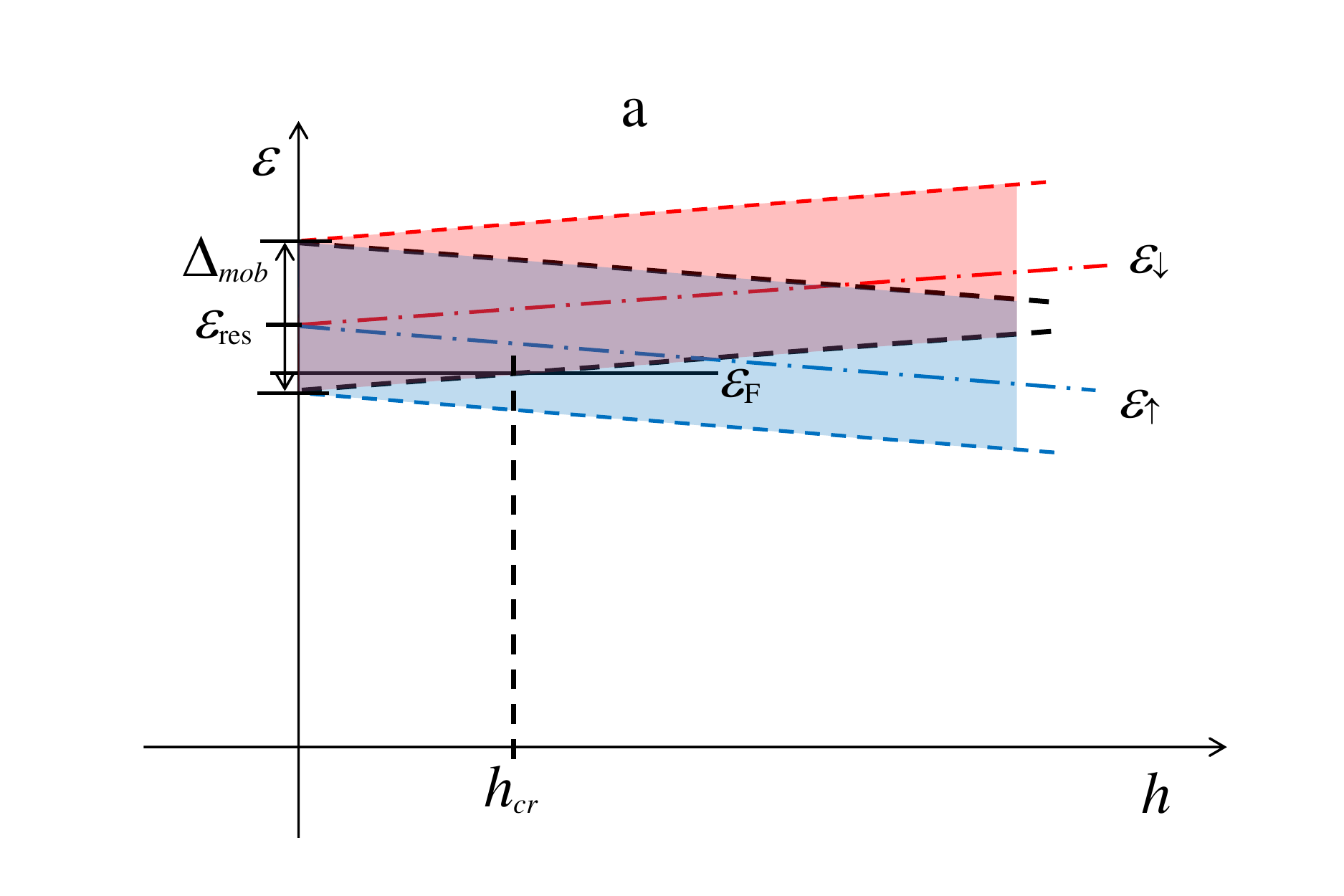}
\includegraphics[width=8cm]{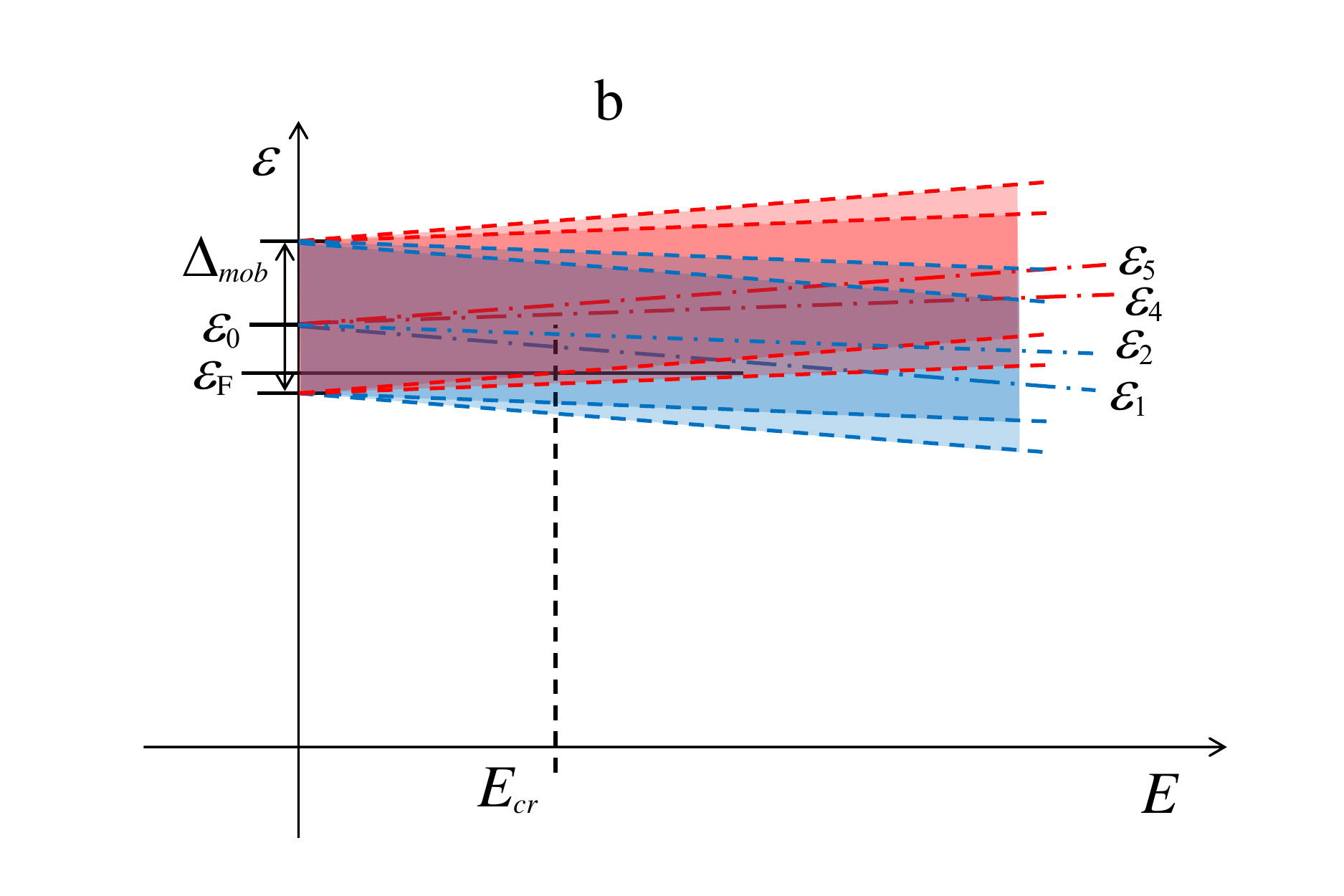}
\caption{Realization of MIT in AGNR with Cu top impurities by tuning of the composed mobility edges: 
a) with an applied magnetic field or b) with an applied electric field (here for $M = 5$).}
\label{fig10}
\end{figure}

But a true electronic phase transition at zero temperature can be realized, for instance, applying a uniform 
static magnetic field $\bf h$ to AGNR. This will produce a spin splitting of the Dirac-like subbands 
defined in Sec. \ref{GF} and also of the impurity levels, implying respective splitting of IRM 
critical points for spin subbands. 

At the same time, the position of overall Fermi level for a given impurity concentration $c$ will 
stay the same as it was for $h = 0$, due to the persisting constancy of the overall DOS. Then, in 
the situation of several overlapping subbands, the overall mobility edges are determined by the Mott 
principle: if, at a given energy, there is at least one state extended, all other states at this 
energy are also extended. Then the overall mobility gap is formed by the intersection of partial 
(formal) gaps for each spin projection and it gets reduced with growing splitting $\m_{\rm B} h$. 
In this way, the overall mobility edges are tuned by the applied field and MIT is realized at its 
critical value $h_{cr} \sim |\e_{\rm F} - \e_g|/\m_{\rm B}$ (see Fig. \ref{fig10}a). But for the 
relevant energy scales of several meV, this may require high enough $h$ values of several tens Tesla. 

An alternative way may be sought in applying a static electric field $E_y \equiv E$ across the 
nanoribbon (along the $y$-axis in Figs. \ref{fig1}, \ref{fig4}) to produce linearly growing local 
potentials $V_m = m e E$  on $m$-layers. This can be shown not to influence the relevant Dirac-like 
subbands, however to produce an $M$-fold splitting between the local energy levels for impurities on 
different $m$-layers and so between the respective mobility gaps, with a subsequent decrease of the 
resulting mobility gap (as in Fig. \ref{fig10}b). The critical fields to achieve MIT in this case, 
$E_{cr} \sim |\e_{\rm F} - \e_g|/(eM)$ of some mV/nm, could be reached with no experimental 
difficulties.

\section{Spin-orbit effects on electronic phase states}\lb{SO}

Yet one more tuning mechanism can result from the spin-orbit (SO) effect, including the Rashba 
spin-orbit coupling \cite{Bychkov}. The latter is generally known to lift the spin degeneracy in the 
systems with broken mirror symmetry, for instance, under electric field $E_z$, applied normally to the 
crystal surface \cite{Vas'ko} or to the 2D graphene plane (along the $z$-axis in Fig. \ref{fig4}) 
\cite{Kane, Kuemmeth}. A similar effect can be achieved in graphene nanotubes \cite{Huertas-Hernando, Rudner} 
and also in AGNRs \cite{Lenz}. 
\begin{figure}[h]
\includegraphics[width=8cm]{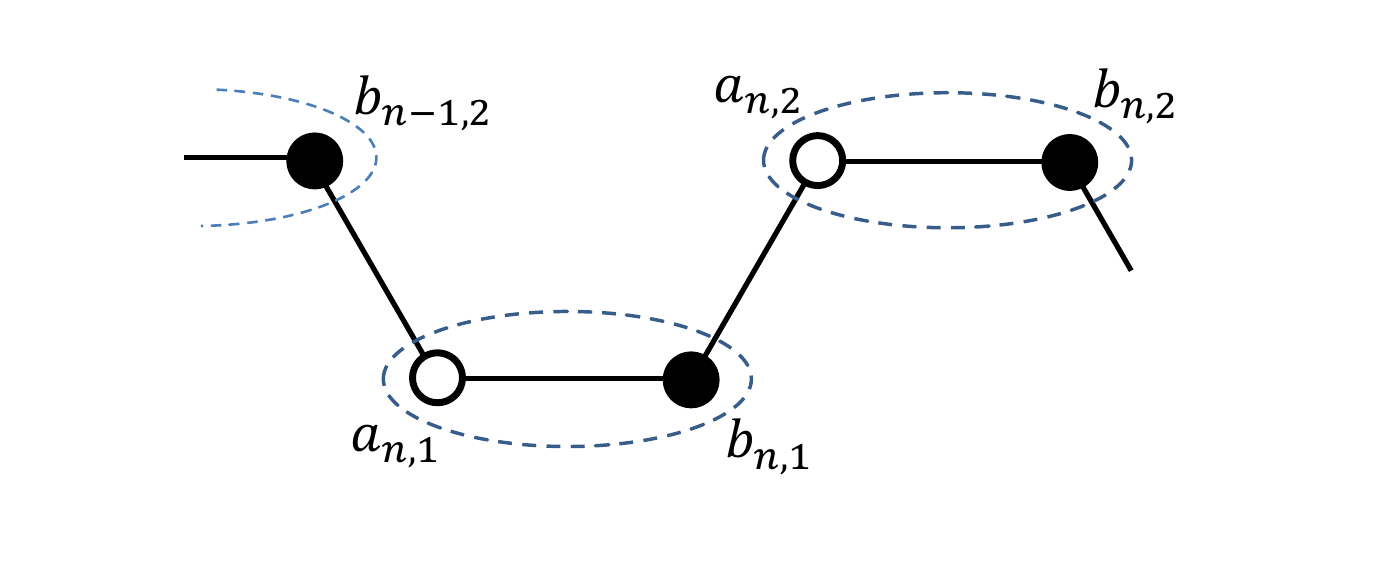}
\caption{Coupled pair of $m = 1,2$ layers in AGNR.}
\label{segm12}
\end{figure}

For the AGNR case, we note that the relevant Dirac-like $\psi_{\pm,k}$ modes (from Eq. \ref{ArmL}) 
have their amplitudes $\sin m\tfrac\pi3$ equal zero at each third $m$-layer, so the interlayer 
couplings through these modes are restricted to the $m$-pairs: $(1,2), (4,5),\dots(M-1,M)$, and 
their overall effect on AGNR can be represented by a single pair, for instance, with $m = 1,2$ 
(see Fig. \ref{segm12}). All the couplings in such a pair of layers are suitably presented in terms 
of local operators, now equipped with explicit $\uparrow\downarrow$ spin indices and then composed 
into 4-spinors: 
\be
f_{n,m} = \left(\begin{array}{c}
    a_{n,m,\uparrow} \\
    a_{n,m,\downarrow}\\
    b_{n,m,\uparrow}\\
    b_{n,m,\downarrow}\\
\end{array}\right).
\lb{ls}
\ee
In the basis of these local spinors, the SO Hamiltonian reads:
\bea
H_{SO} & = & \sum_n \left[f_{n,1}^\dagger\hat H_{SO}\left(f_{n,1} + f_{n,2} + f_{n - 1,2}\right)
\right.\nn\\
& + & \left.f_{n,2}^\dagger\hat H_{SO}\left(f_{n,2} + f_{n,1} + f_{n + 1,1}\right)\right],
\eea
where the 4$\times$4 matrix:
\[\hat H_{SO} = \D\hat\s_z + \l\left(\hat\s_x\hat\t_y - \hat\s_y\hat\t_x\right).\]
includes the Pauli matrices $\hat\s_j$ in spin $\uparrow,\downarrow$-indices and $\hat\t_j$ in 
sublattice $a,b$-indices and also the parameters $\D$ for standard SO and $\l$ for Rashba SO (the 
latter being $E_z$-dependent). The estimates for these local SO couplings in 2D graphene (also 
plausible for nanoribbons) show the standard $\D \sim 10^{-4}$ \cite{Kane}, fixed and much 
smaller of the relevant energy scales for MIT crossing. Otherwise, the Rashba $\l$ can be strongly 
enhanced \cite{Marchenko} and yet tunable \cite{Kane}, so it is taken as an effective SO variable 
below. 

Next, we pass to the basis of chain-wave 4-spinors:
\be 
\psi_k = \left(\begin{array}{c}
    \psi_{+,k,\uparrow} \\
    \psi_{+,k,\downarrow}\\
    \psi_{-,k,\uparrow}\\
    \psi_{-,k\downarrow}\\
\end{array}\right),
\lb{cws}
\ee
which are related to the local spinors by Eq. \ref{ls} (with $m = 1,2$) through a $\hat\t$-rotation:
\be
f_{n,m} = \frac {(-1)^{m - 1}}{2\sqrt{\nu N}}\sum_k {\rm e}^{ik\xi_{n,m}}\hat U_k\psi_k.
\ee
Here the rotation matrix:
\be
\hat U_k = \cos\phi_k\left(\hat\t_x - \hat\t_z\right) + \sin\phi_k\left(\hat\t_y + i\hat\t_0\right),
\ee
results from Eq. \ref{rel1} (restricted to single $j = 2\nu$) for the components of  $f_{n,m}$. At
relevant $|k| \ll 1$, the phase $\phi_k \equiv \phi_{2f,k}$ is approximated as $\phi_k \approx \phi_0 
= -\pi/4$. Then the Dirac-like part of SO Hamiltonian reads:
\be
H_{SO} = \sum_k \psi_k^\dagger\left[-\D\hat\s_z\hat\t_x + \l\left(\hat\s_x\hat\t_z + \hat\s_y\hat
\t_y\right)\right]\psi_k,
\ee
and, together with the Dirac-like part of $H_{tb}$ by Eq. \ref{Hd}, it defines the SO-split dispersion 
laws (in neglect of impurity disorder):
\be 
\e_{\pm,k} = \sqrt{\e_k^2 + \D_\pm^2}.
\lb{dl}
\ee
\begin{figure}[h]
\includegraphics[width=7cm]{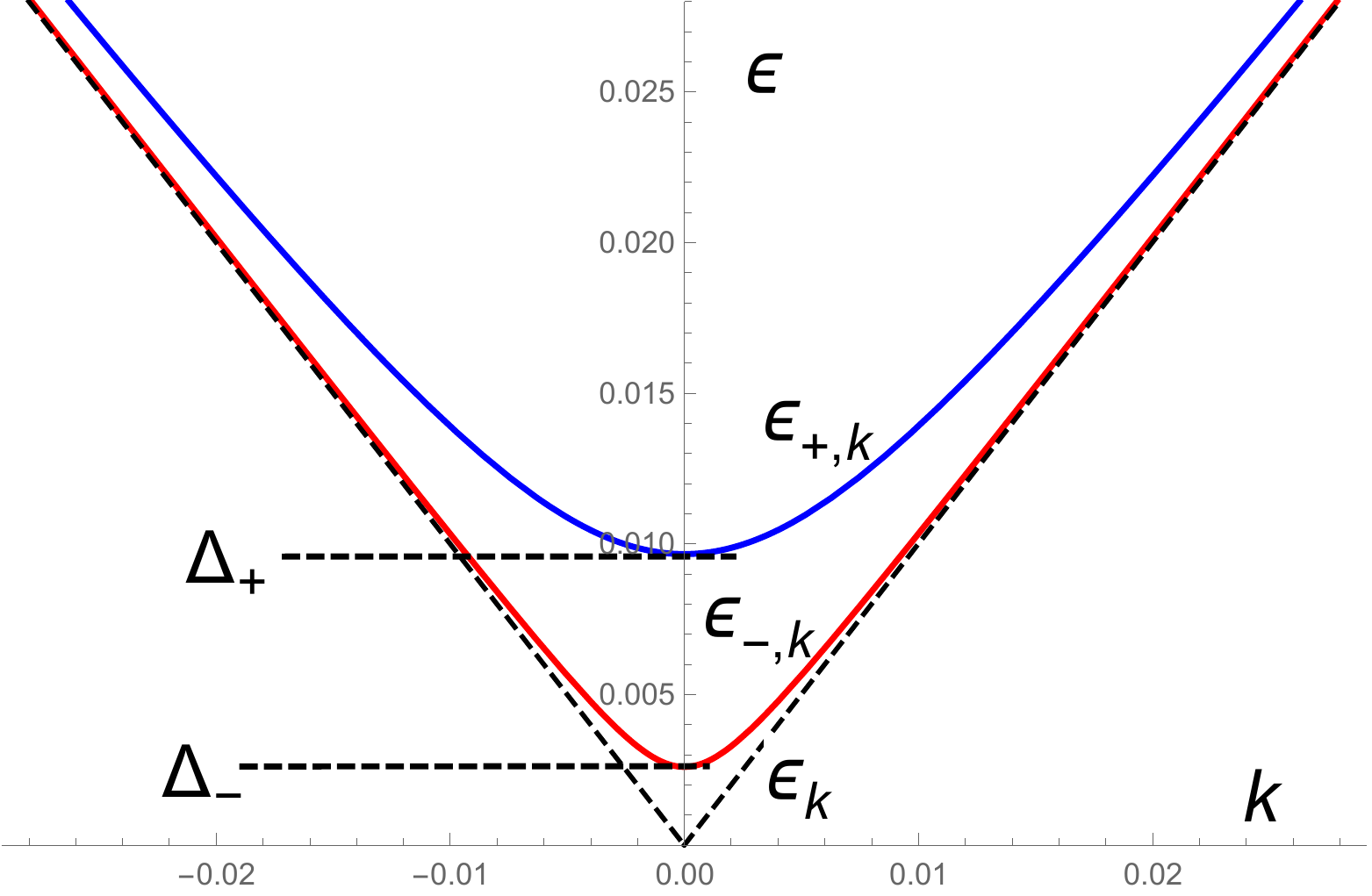}
\caption{SO splitting of the Dirac-like mode $\e_k$ (dashed lines) into $\e_{\pm,k}$ at the choice of 
$\l = 5\cdot10^{-3}$ and $\D = 2.5\cdot10^{-4}$.}
\label{fig11}
\end{figure}
Here the non-zero bandgaps:
\be 
\D_\pm = \sqrt{\D^2 + \l^2\left(2 \pm \sqrt{3}\right)},
\lb{bg}
\ee
 are due to both SO types but their splitting is only due to the Rashba SO as shown in Fig. \ref{fig11} 
for the choice of SO parameters $\D = 2.5\cdot10^{-4}$ \cite{Kane} and $\l = 5\cdot10^{-3}$ \cite{Marchenko}, 
and they create the low-energy DOS singularities:
\begin{figure}[h]
\includegraphics[width=7cm]{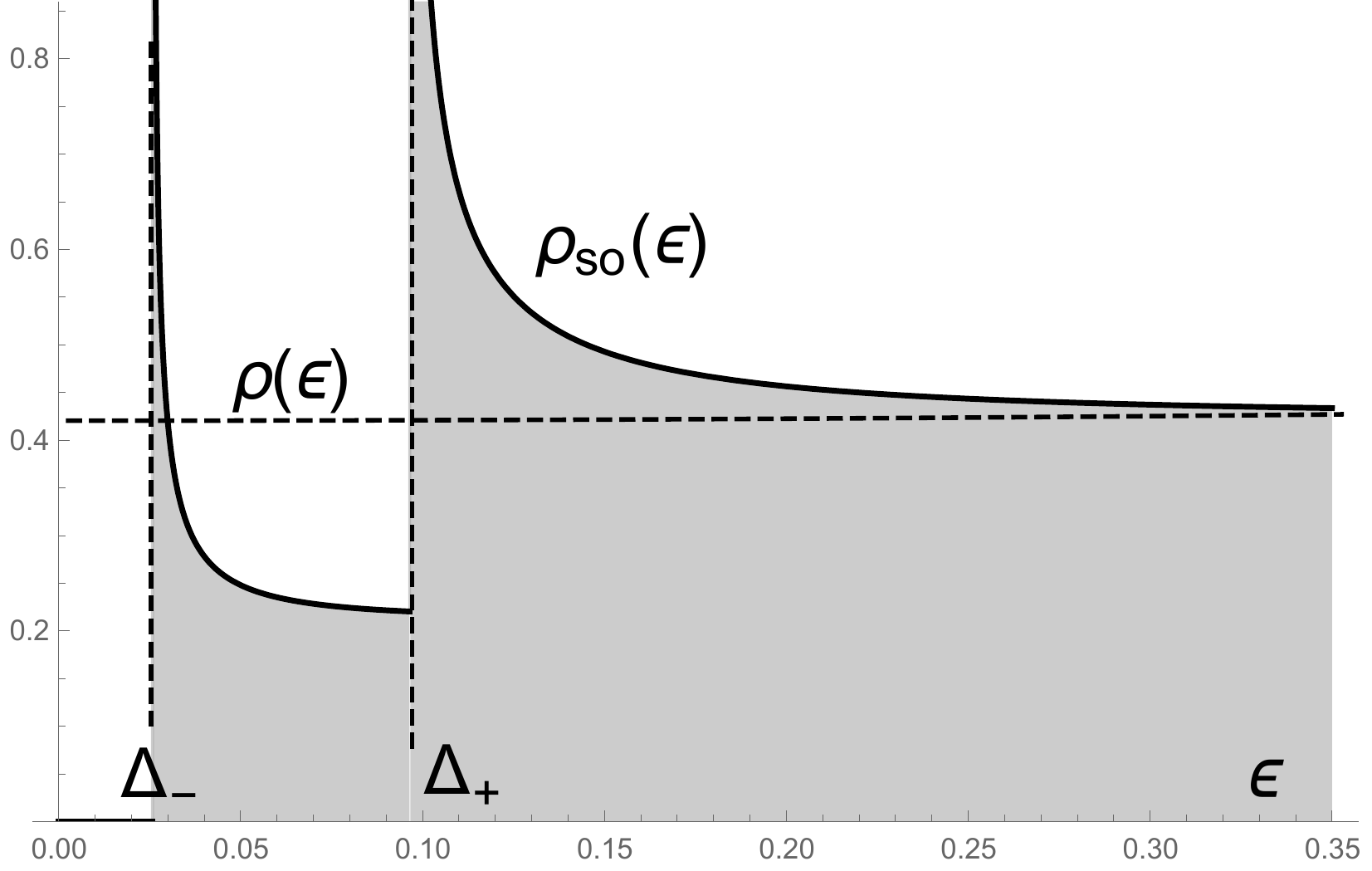}
\caption{Singularities of the low-energy DOS of 5-AGNR with split subbands at the same choice of SO 
parameters as in Fig. \ref{fig11}.}
\label{soDOS}
\end{figure}
\be
\r_{so}(\e) = \frac{\e\r_0}2\left[\frac{\theta(\e - \D_+)}{\sqrt{\e^2 - \D_+^2}} + \frac{\theta(\e - \D_-)}
{\sqrt{\e^2 - \D_-^2}}\right],
\ee
as shown in Fig. \ref{soDOS}. 

\begin{figure}[h]
\includegraphics[width=8cm]{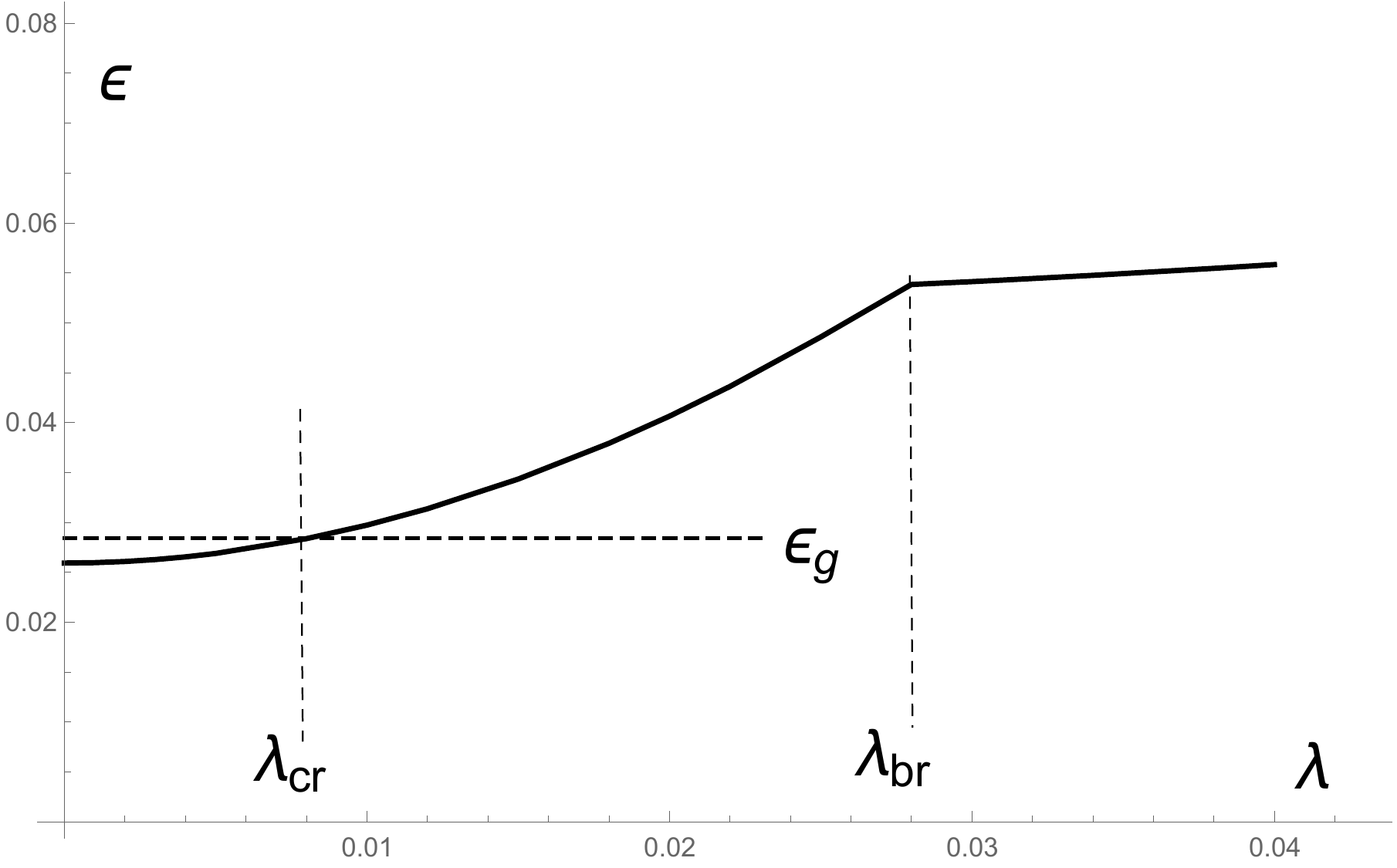}
\caption{Dependence of the Fermi level on Rashba SO coupling for 5-AGNR with Cu top impurities 
concentration fixed at $c = 0.011$, its crossing with the mobility edge $\e_g$ at $\l = \l_{cr} \approx 
0.0078$ indicates an SO-tuned MIT to occur.}
\label{fig12}
\end{figure}

Such DOS behavior, instead of its almost constancy at no account of SO (by Eq. \ref{rd}), when used 
in Eq. \ref{ef} leads to the $\l$-dependence of the Fermi level given by the equation:
\be
\sqrt{\e_{\rm F}^2 - \D_+^2} + \sqrt{\e_{\rm F}^2 - \D_-^2} \approx 2c/\r_0.
\lb{eFl}
\ee
Its numerical solution defines MIT to occur when $\l$ reaches a certain critical value $\l_{cr}$ as shown 
in Fig. \ref{fig12} . 

Noting that for all relevant impurity concentrations $c > c_0$ we have $c/\r_0 \gg \D$, the approximate 
solution of Eq. \ref{eFl} for $\l \ll c/\r(0)$ reads: 
\be
\e_{\rm F}(\l) \approx \frac c{\r_0} + \frac{\r_0}c\left(\D^2 + 2\l^2\right).
\ee

This quadratic $\l$-dependence relates to filling of both $\e_{\pm,k}$ subbands by $c$ charge 
carriers, it follows the lowest part of the numerical solution in Fig. \ref{fig12}. With further growing 
$\l$, a break of $\e_{\rm F}(\l)$ occurs when it gets intercepted by the faster growing upper 
bandgap $\e_+$ at the break value $\l_{br} = c/(\sqrt{2\sqrt3}\r_0)$. The next slower $\e_{\rm F}
(\l)$ growth at $\l > \l_{br}$ relates to filling of only the lower subband which is expressed as:
\be
\e_{\rm F}(\l) = \sqrt{(c/\r_0)^2 + \D^2 + (2 - \sqrt3)\l^2}.
\ee

Notably, for the considered case of Cu top impurities, both the impurity resonance $\e_{res}$ and 
the mobility edge $\e_g$ lie in the energy range $\e \gg \e_\pm$ where $\r_{SO}(\e)$ already reaches 
its asymptote $\approx \r(0)$, so the mobility gap structure stays practically insensitive to Rashba SO. 
Hence a possibility arises here for MIT to be realized by SO tuning of $\e_{\rm F}$ at fixed $\e_g$, 
unlike the above considered regimes with tuning of mobility edges at fixed Fermi level. 

Notably, this tuning process can be realized in a combined way: a rough "tuning" of $\e_{\rm F}$ 
closeness to $\e_g$ by a proper choice of impurity parameters $\e_{res},\,\o,\,c$ and also by a 
strong structural contribution to the Rashba parameter $\l$, say, from a gold substrate atomic field 
\cite{Marchenko}, and then its fine tuning by an applied external field. Evidently, the expected MIT 
at such sub-meV energy scales would require a range of liquid He temperatures for its sufficient 
resolution.

\section{Discussion}\lb{Dis}

The obtained results demonstrate how the difference of electronic states in graphene nanoribbons 
defined by their edge orientations is reflected in their stability against impurity disorder. Physically, 
this opens the possibility for specific electronic phase transitions and for their controls by combining 
the disorder and external bias effects. 

It is of interest to compare the above picture of 1D spectrum reconstruction under AM impurity disorder 
with the known such effects in 3D and 2D systems. This comparison can be done between the corresponding 
correlators, defined by the system dimensionality and low energy quasiparticles dispersion. For instance, 
such a correlator in 3D semiconductors (quadratic dispersion) \cite{Ivanov1979},  decays with distance $r$ as:
$\propto (\sin \sqrt{\e}r)/(\sqrt{\e} r)$, a similar behavior for 3D acoustic phonons (linear dispersion) 
\cite{Ivanov1970} is found as: $\propto (\sin \e r)/(\e r)$, and it is modified to $ \propto (\sin \e r)/\sqrt{\e r}$ 
in 2D graphene with linear Dirac dispersion \cite{Pogorelov2020}. All these systems admit both extended and 
localized quasiparticles based on impurity states \cite{ILP}. But, unlike all those, a constancy or only a weak 
exponential decay of correlators are found for the present AGNR case (see Eqs. \ref{cor0}, \ref{cor3} below) 
which defines the complete localization of all the states near $\e_{res}$ (as in Fig. \ref{fig6}a, unlike the  
2D graphene case in Fig. \ref{fig6}b). Another AGNR specifics, the low-energy DOS constancy, defines a 
higher sensitivity of Fermi level to doping and so more possibilities for tuning of the system electronic 
properties.

The present study was limited to the simplest framework of tight-binding model for pure nanoribbons 
and simplest models for impurity perturbations on them. In principle, it can be extended to account for 
many other physical factors, as electron-electron Hubbard correlations, spin-ordering effects, phonon 
and spinon excitations, etc.

Also, the effects from passivating hydrogens, known to be commonly present at the edges of experimental 
nanoribbon samples \cite{Wang2016,Ruffieux2016,Wang2017,El Abbassi}, may influence the dynamics of host 
nanoribbon carriers. This factor can be naturally included into the above developed Hamiltonians and 
resulting GFs, to be possibly an object of future study. At least, an experimental check for the suggested 
effects, for instance, on carriers mobility and its collapse under definite external factors should be 
of considerable interest.

\section*{Acknowledgments}

We are grateful to Larissa S. Brizhik, Mark I. Dykman and Aleksandr A. Eremko for useful discussions of this work.
The work by V.M.L. was partially supported by Grant Nos. 0117U000236 and 0117U000240 from the Department 
of Physics and Astronomy of the National Academy of Sciences of Ukraine.

\appendix
\section{Beyond the T-matrix}

Validity of the above T-matrix solution should be yet verified in view of the effective 1D character 
of the relevant Dirac-like quasiparticles. It is known that generic 1D systems are unstable against 
any disorder, up to a full localization of all their eigen-states \cite{Lifshitz}, and the IRM check 
with use of the simplest (single-impurity) T-matrix is not sufficient to detect this. Therefore the 
IRM results from Sec. \ref{GF} need a support by some T-matrix extensions known for many 
other disordered systems. There are two such possible extensions: 

1) group expansions (GEs) in clusters of correlated impurity centers \cite{ILP} and 

2) self-consistent T-matrix approximation \cite{Elliott}.

For GEs, their basic elements are the correlators, defined for different GE forms. The simplest 
form is that of non-renormalized GE, known to better apply for the energy ranges of localized states. Here 
the correlator for the considered Dirac-like quasiparticles is written as:  
\be
A_r(\e) = \frac{2T(\e)}{3\nu N}\sum_k {\rm e}^{ikr}\left(\frac1{\e - \e_k} + \frac1{\e + \e_k}\right)
\lb{cor}
\ee
(taking into account equal contributions from $j = \nu$ and $j = 2\nu$ modes). Then, after integration 
in $k$ by Eq. \ref{int}, the related integral has its long distance asymptotics at $r \gg 1$ as:
\be
A_r(\e) =\frac{2T(\e)\e}{3\pi \nu}\int_{-\pi}^{\pi}\frac{{\rm e}^{ikr}dk}{\e^2 - 4\sin^2 k/2} \approx 
\frac{2T(\e)\e}{3\nu}\sin\e r,
\lb{cor0}
\ee
that is non-decaying. This contrasts with the decaying correlators in 3D and 2D systems and makes 
all the GE terms for the 1D-like system formally divergent. To avoid that problem, some alternative, 
renormalized GE forms (more adequate for conductive states) could be employed.

For instance, the first order renormalization for GE is obtained with the simple change, $\e \to \e 
- cT(\e)$ in the denominators of Eq. \ref{cor}. Then the renormalized correlator $\tilde A_r(\e) = 
2T(\e)[\e - cT(\e)]\tilde I_r(\e)/(3\nu)$, involves the integral:
\be
\tilde I_r(\e) = \frac 1{2\pi}\int_{-\pi}^\pi \frac{{\rm e}^{ikr}dk}{[\e - cT(\e)]^2 
- 4\sin^2 k/2}.
\lb{cor1}
\ee
\begin{figure}
\includegraphics[width=8cm]{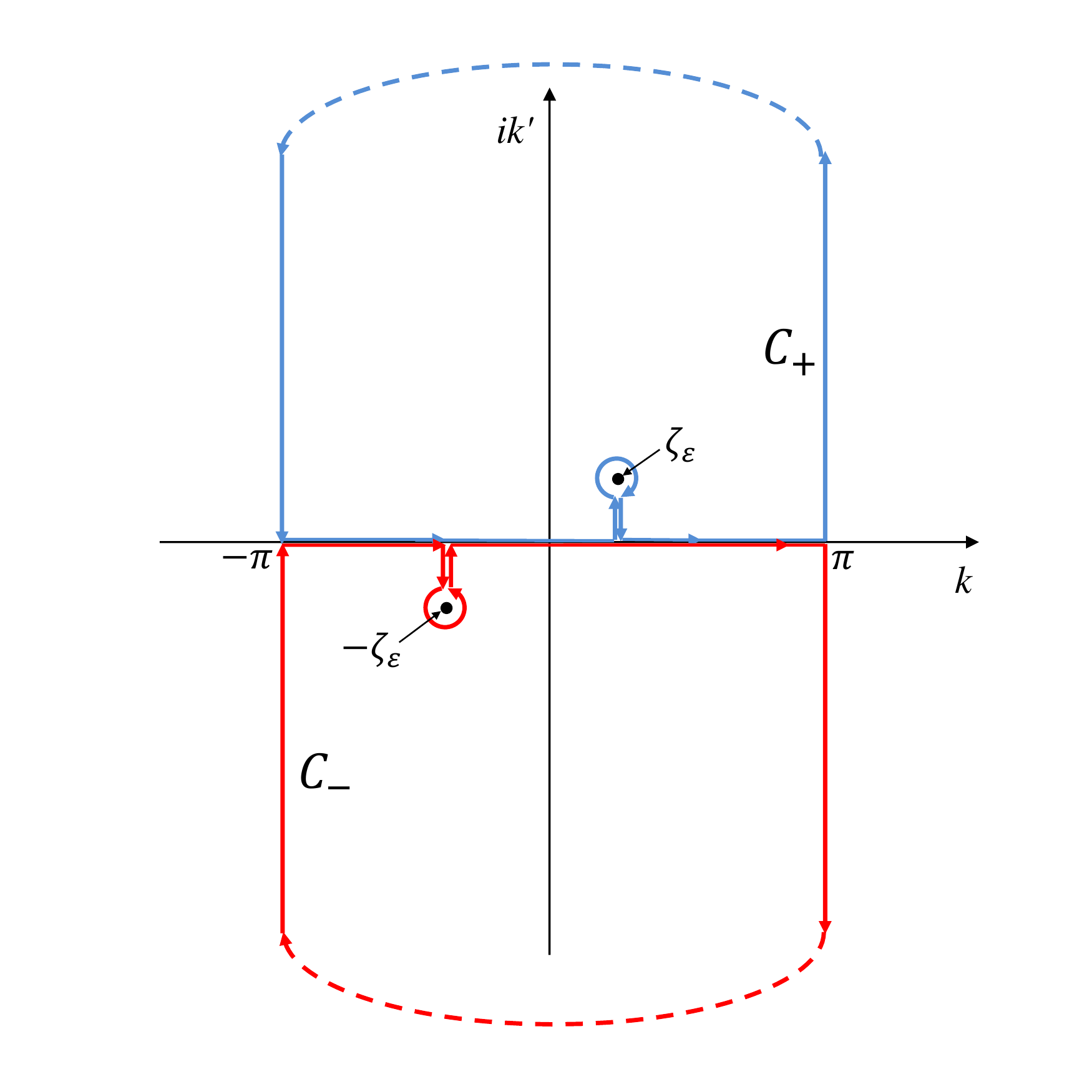}
\caption{Integration contours for calculation of the renormalized correlator $\tilde A_r(\e)$, Eq. 
\ref{cor1}: $C_+$ (blue lines) for $r > 0$ and $C_-$ (red lines) for $r < 0$.}
\label{fig8}
\end{figure}
This can be found analytically, passing to complex momentum: $k \to \z = k + ik'$ and extending 
integration to one of the closed contours shown in Fig. 8, depending on the 1D correlator direction. 
The forward direction, $r > 0$, relates to $C_+$ (blue lines) with the pole $\z_\e = 2\arcsin [\e - 
cT(\e)]/2$, and the backward direction, $r < 0$, does to $C_-$ (red lines) with the pole $-\z_\e$.

The contour integral for the forward case:
\be
\frac 1{2\pi}\oint_{C_+} \frac{{\rm e}^{i\z r}}{\e - cT(\e) - 2\sin \z/2}d\xi,
\lb{cont}
\ee
 presents a zero sum of three terms: $0 = I_r(\e) + R_r(\e) + V_r(\e)$. Here the residue term, 
 $R_r(\e)  = {\rm e}^{i\z_\e r}/\cos (\z_\e/2)$, and the term from the semi-infinite vertical 
 segments, $V_r(\e) \approx  (2i\sin\pi r)/[(1 - \e)r]$, define the sought correlator as:
 \bea
&&\tilde A_r(\e) = \frac{2T(\e)[\e - cT(\e)]}{3f}\nn\\
&&\qquad\qquad\times\,\left[\frac{{\rm e}^{i\z_\e r}}{\cos (\z_\e/2)} + 2i\frac{\sin\pi r}
{(1 - \e)r}\right).
\lb{cor2}
\eea
For the low energy range, $|\e|, c|T(\e)| \ll 1$ and $\z_\e \approx \e - cT(\e)$, Eq. \ref{cor2} 
simplifies to:
\bea
&&\tilde A_r(\e) \approx \frac{2T(\e)[\e - cT(\e)]}{3\nu}\nn\\
&& \times\,\left[{\rm e}^{i[\e - c {\rm Re}\,T(\e)]r}{\rm e}^{-c|{\rm Im}\,T(\e)r|} + 2i\frac{\sin\pi 
r}r\right].
\lb{cor3}
\eea
Here, unlike Eq. \ref{cor0} both terms in the brackets are already decaying with $r$. For the 
backward case, integration over $C_-$ gives the same result.

The relevant criterion for GE convergence, the smallness of the dominating contribution by impurity 
pair clusters into the quasiparticle self-energy compared to that by single impurities \cite{ILP}, is 
presented here in the form:
\be
B_2(\e) \approx c\left|\int_0^\infty  \tilde A_{-r}(\e)\tilde A_r(\e)^2{\rm e}^{-i(\e - c{\rm Re}\,
T(\e))r}dr\right| \ll 1.
\lb{conv}
\ee
\begin{figure}[h]
\includegraphics[width=8cm]{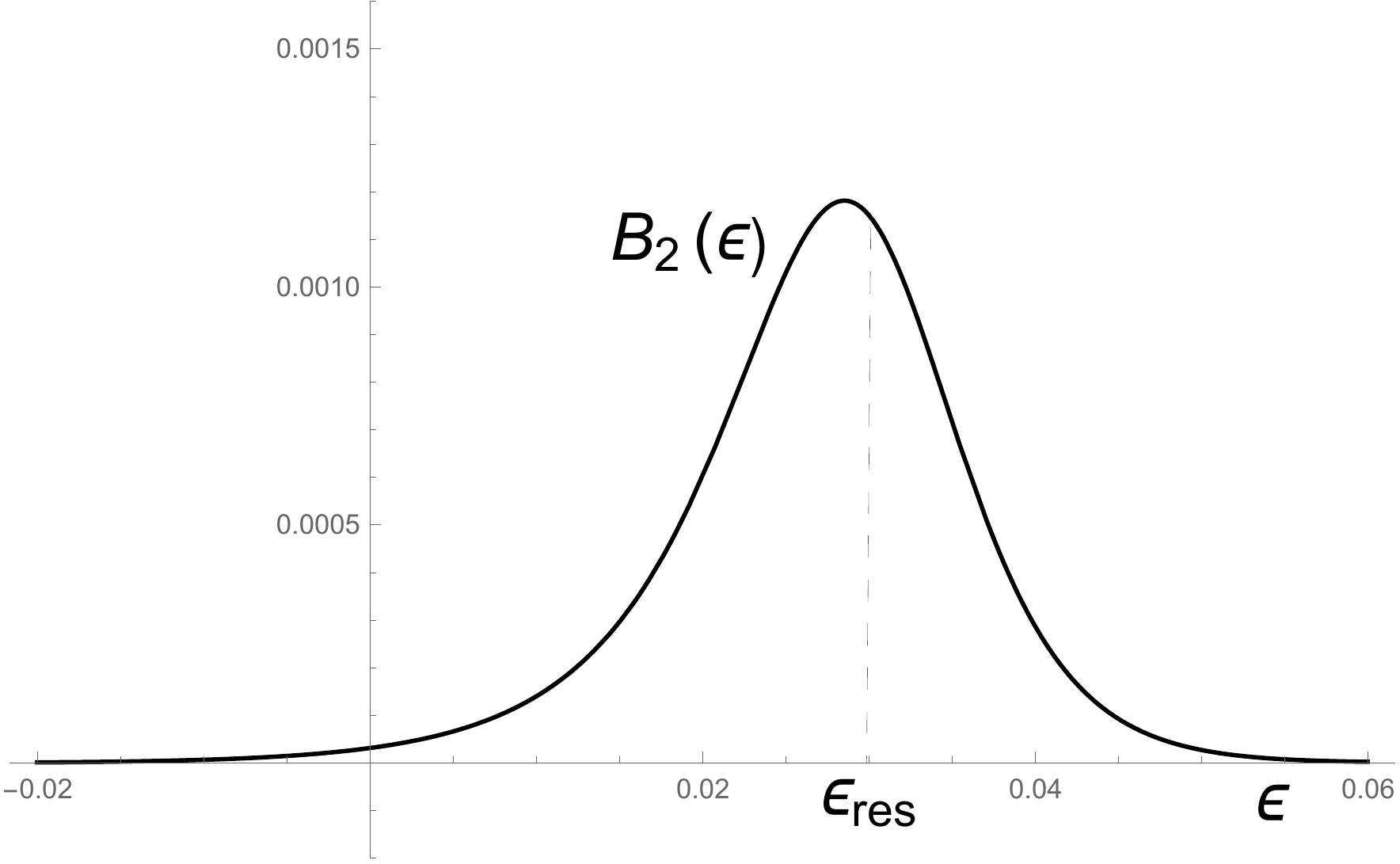}
\caption{Smallness of the relative contribution to GE by impurity pairs, $B_2(\e)$ by Eq. \ref{conv1} 
(for Cu impurities with concentration $c = 0.05$ in 5-AGNR), assuring GE convergence for this system.}
\label{fig9}
\end{figure}
Then the residue term in Eq. \ref{cor3} with its slower exponential decay $\propto {\rm e}^{-c|
{\rm Im}\,T(\e) r|}$ dominates in the $r$-integral and converts the above criterion into:
\be
B_2(\e) \approx \frac8{81\nu^3}\left|\frac{T^3(\e)[\e - cT(\e)]^3}{{\rm Im}\,T(\e)}\right| 
\ll 1.
\lb{conv1}
\ee
The straightforward numerical check shows this criterion to surely hold for all the above 
considered impurity parameters (see an  example in Fig. \ref{fig9}). 

Also, the self-consistent extension of the T-matrix function:
\be
T_{sc}(\e) = \frac{\o^2}2\left\{\e - \e_{res} - \frac{i\o^2}{4\nu\sqrt{1 - \left[\frac{\e - cT_{sc}
(\e)}2\right]^2}}\right\},
\ee
practically coincides with its non-renormalized version $T(\e)$ by Eq. \ref{TM1}, due to the 
above noted smallness of $|\e - cT(\e)| \ll 1$, assured for all $|\e| \ll 1$.

Hence, the discussed T-matrix results for the quasiparticle spectra in disordered AGNRs and the 
related estimates for their mobility edges can be considered valid. We conclude that the inverse 
lifetime by the non-renormalized T-matrix in r.h.s. of Eq. \ref{IRM1} is the main factor for 
quasiparticle localization in AGNRs. So the results of Sec. \ref{GF} correctly determine the 
system observable characteristics.


\begin{thebibliography}{1}
 \bibitem{Geim2004} K.S. Novoselov, A.K. Geim, S.V. Morozov, S.V. Dubonos, Y. Zhang, D. Jiang, arXiv:cond-mat/0410631 (2004).
 \bibitem{Geim2005} S.V. Morozov, K.S. Novoselov, F. Schedin, D. Jiang, A.A. Firsov, A.K. Geim, Phys. Rev. B, 72, 201401 (2005).
 \bibitem{Geim2009}  A.H. Castro Neto, F. Guinea, N.M.R. Peres, K.S. Novoselov, A.K. Geim, Rev. Mod. Phys., 109, 109 (2009).
\bibitem{Yang} L. Yang, C.-H. Park, Y.-W. Son, M.L. Cohen, S.G. Louie, Phys. Rev. Lett., 99, 186801 (2007).
\bibitem{Yang2008} X. Yang, X. Dou, A. Rouhanipour, L. Zhi, H.J. Räder, K. M\"ullen, J. Amer. Chem. Soc., 130, 4216(2008).
\bibitem{Ruffieux}  P. Ruffieux, J. Cai, N.C. Plumb, L. Pattney, D. Prezzi, A. Ferretti, E. Molinari, X. Feng, K. M\"ullen, C.A. Pignedoli, 
R. Fasel, ACS Nano, 6, 6930 (2012).
\bibitem{Wakabayashi1996}  M. Fujita, K. Wakabayashi, K. Nakada, K. Kusakabe, J. Phys. Soc. Jpn, 65, 1920 (1996).
\bibitem{Pogorelov2021} Y.G. Pogorelov, D. Kochan, V.M. Loktev, Low Temp. Phys., 47, 754 (2021).
\bibitem{Wakabayashi2009} K. Wakabayashi, Y. Takane, M. Yamamoto, M. Sigrist, New Journal of Physics, 11, 095016 (2009).	
\bibitem{Wakabayashi2010}   K. Wakabayashi, K. Sasaki, T. Nakanishi, T. Enoki, Sci. Technol. Adv. Mater., 11, 054504 (2010).
\bibitem{Zubarev1960} D. Zubarev, Sov. Phys. Uspekhi, 3, 320 (1960).
\bibitem{Economou1979}  E.N. Economou, Green's Functions in Quantum Physics, Springer (1979).
\bibitem{Lifshitz_1963} I. M. Lifshitz, Soviet Physics JETP, 17, 1159 (1963).
\bibitem{Anderson} P. W. Anderson, Phys. Rev., 124, 41 (1961).
\bibitem{Bonch} V.L. Bonch-Bruevich and S.V. Tyablikov, The Green Function Method in Statistical Mechanics, North Holland Publishing 
House (1962).
\bibitem{Pogorelov2020}  Y.G. Pogorelov, V.M. Loktev, D. Kochan, Phys. Rev. B, 102, 155414 (2020).
\bibitem{IoffeRegel} A.F. Ioffe and A.R Regel, Progress Semicond., 2, 237 (1960).
\bibitem{Mott}  N.F. Mott, Phil. Mag., 102, 835 (1962).
\bibitem{Bychkov}  Yu. A. Bychkov and E. I. Rashba, Sov. Phys. - JETP Lett., 39, 78 (1984).
\bibitem{Vas'ko} F. T. Vas'ko, Sov. Phys. - JETP Lett., 30, 541 (1979).
\bibitem{Kane} C.L. Kane and E.J. Mele, Phys. Rev. Lett., 95, 226801 (2004).
\bibitem{Kuemmeth} Ferdinand Kuemmeth and Emmanuel I. Rashba, Phys. Rev. B, 80, 241409 (2009).
\bibitem{Huertas-Hernando} D. Huertas-Hernando, F. Guinea and A. Braatas, Phys. Rev. B, 74, 155426 (2006).
\bibitem{Rudner} Mark S. Rudner and Emmanuel I. Rashba, Phys. Rev. B, 81, 125426 (2010).
\bibitem{Lenz} Lucia Lenz, Daniel F. Urban and Dario Bercioux, Eur. Phys. J B, 86, 502 (2013).
\bibitem{Marchenko} D. Marchenko, A. Varykhalov, M.R. Scholz, G. Bihlmayer, E.I. Rashba, A.M. Shikin and O. Rader, Nat. Commun, 
3, 1 (2012).
\bibitem{Wang2016} S. Wang, L. Talirz, C.A. Pignedoli, X. Feng, K. M\"ullen, V. Meunier, R. Fasel and P. Ruffieux, Nat. Commun., 
7, 11507 (2016).
\bibitem{Ruffieux2016} P. Ruffieux, S. Wang, B. Yang, C. Sanchez-S\'anchez, J. Liu, T. Dienel, L. Talirz, P. Shinde, C.A. Pignedoli, 
D. Passerone, T. Dumslaff, X. Feng, K. M\"ullen and R. Fasel, Nature, 531, 489 (2016).
\bibitem{Wang2017} S. Wang, N. Kharche, E. Costa {Gir\~ao}, X. Feng, K. M\"ullen, V. Meunier, R. Fasel, P. Ruffieux, Nano Lett., 17, 
4277 (2017).
\bibitem{El Abbassi} M. El Abbassi, M.L. Perrin, G. Borin Barin, S. Sangtarash, J. Overbeck, O. Braun, C.J. Lambert, Q. Sun, T. Prechtl, 
A. Narita, K. M\"ullen and P. Ruffieux, H. Sadeghi, R. Fasel and M. Calame, ACS Nano, 14, 5754 (2020).
\bibitem{Lifshitz} I. M. Lifshits, S. A. Gredeskul, L. A. Pastur, Introduction to the theory of disordered systems, Wiley, New York (1988).
\bibitem{ILP} M.A. Ivanov , V.M. Loktev and Yu.G. Pogorelov, Physics Reports, 153, 209 (1987).
\bibitem{Elliott} R. J. Elliott, J. A. Krumhansl and P. L. Leath, Rev. Mod. Phys., 46, 465 (1974).
\bibitem{Ivanov1979} M.A. Ivanov and Yu.G. Pogorelov, Sov. Phys. JETP, 79, 1010 (1979).
\bibitem{Ivanov1970} M.A. Ivanov, Sov. Phys. Solid State, 12, 1508 (1970).

\end{thebibliography}
\end{document}